\documentclass[final,1p,times]{elsarticle}

\usepackage{amsmath}
\usepackage{amssymb}
\usepackage{amsfonts}
\usepackage{float}
\usepackage{graphicx}
\usepackage{tabularx}
\usepackage{multirow}
\usepackage{lscape}
\usepackage{rotating}
\usepackage{pstricks}

\newcommand{\bk}{\boldsymbol{\mathbf k}}
\newcommand{\bkp}{\boldsymbol{\mathbf k^\prime}}

\def\be{\begin{equation}}
\def\ee{\end{equation}}
\def\bea{\begin{eqnarray}}
\def\eea{\end{eqnarray}}
\def\bean{\begin{eqnarray*}}
\def\eean{\end{eqnarray*}}
\def\nn{\nonumber}
\def\nnn{\nonumber \\}

\journal{Annals Physics}

\begin{document}

%%%%%%%%%%%%%%%%%%%%%%%%%%%%%%%%%%%%%%%%%%%%%%%%%%%%%%%%

\begin{frontmatter}

\title{Infinite matter properties and zero-range limit of non-relativistic finite-range interactions}

\author[1]{D. Davesne} 
\author[1]{P. Becker}
\author[2]{A. Pastore}
 \author[3]{J. Navarro}
 
\address[1]{Universit\'e de Lyon, Universit\'e Lyon 1, CNRS/IN2P3, Institut de Physique Nucl{\'e}aire de Lyon, UMR 5822, F-69622 Villeurbanne cedex, France}

\address[2]{Department of Physics, University of York, Heslington, York, Y010 5DD, United Kingdom}

\address[3]{IFIC (CSIC-Universidad de Valencia), Apartado Postal 22085, E-46.071-Valencia, Spain}

\begin{abstract}
We discuss some infinite matter properties of two finite-range interactions widely used for nuclear structure calculations, namely Gogny and  M3Y  interactions. We show that some useful informations can be deduced for the central, tensor and spin-orbit terms from the spin-isospin channels and the partial wave decomposition of the symmetric nuclear matter equation of state. We show in particular that the central part of the Gogny interaction should benefit from the introduction of a third Gaussian and the tensor parameters of both interactions can be deduced from special combinations of partial waves. We also discuss the fact that the spin-orbit of the M3Y interaction is not compatible with local gauge invariance. Finally, we show that the zero-range limit of both families of interactions coincides with the specific form of the zero-range Skyrme interaction extended to higher momentum orders and we emphasize from this analogy its benefits.
\end{abstract}


%%%%%%%%%%%%%%%%%%%%%%%%%%%%%%%%%%%%%%%%%%%%%%%%%%%%%%%%
%    21.30.Fe 	% Forces in hadronic systems and effective interactions
%    21.60.Jz 	% Nuclear Density Functional Theory and extensions 
                % (includes Hartree-Fock and random-phase approximations)
%    21.65.-f 	% Nuclear matter
%    21.65.Mn 	% Equations of state of nuclear matter 
                % (see also 26.60.Kp Equations of state of neutron-star matter)

\end{frontmatter}

%%%%%%%%%%%%%%%%%%%%%%%%%%%%%%%%%%%%%%%%%%%%%%%%%%%%%%%%

\section{Introduction}

The nuclear energy density functional theory (NEDF) is the tool of choice for the description of nuclear properties from drip-line to drip-line and from light to super-heavy elements~\cite{ben03}. Among the different choices for the functional, the most popular are the ones derived from the non-relativistic effective nucleon-nucleon interactions as the zero-range Skyrme~\cite{sky59,vau72}, and the finite-range Gogny~\cite{gog75,dec80} or M3Y~\cite{ber77,nak03} families. 
At the mean field level, the global performances along the nuclear chart are comparable both for ground states~\cite{gor09,gor09b} and excited ones~\cite{ber07,ter08} and there is no clear argument, from this perspective, to prefer a zero-range or a finite-range interaction.

In this article, we aim to further analyse the aforementioned effective interactions by comparing 
their results in infinite nuclear matter. Although the infinite medium is an idealised system, its equation of state, and more particularly its part around saturation density, represents a very important constraint not only for phenomenological interactions~\cite{dut12}, but also for microscopic calculations based on bare nucleon-nucleon interactions. 
We propose here to compare some results obtained from calculations based on the previous phenomenological effective interactions with microscopic calculations based on the bare nucleon-nucleon interaction, as Brueckner-Hartree-Fock (BHF)~\cite{bal97}, Chiral Effective Field Theory ($\chi$-EFT)~\cite{heb11,gez13}, or Many Body Perturbation theory (MBPT)~\cite{bog05,rot08}. These microscopic results represent a very useful input to determine those parts of the effective interactions which are difficult to fix in the standard fitting procedure based on finite nuclei properties.

Thus, besides comparing the symmetric nuclear matter (SNM) equation of state (EoS), we also consider its separate contributions, as its decomposition in spin-isospin $(S,T)$ channels or in partial waves. Comparing such contributions leads to additional constraints providing further insights on those effective interactions terms which do not appear explicitly in the EoS, as the spin-orbit and tensor interaction terms~\cite{Dav15}. One of the goals of this work is to show that it is possible, using a partial wave decomposition of the EoS, to give an alternative view of the main characteristics and limitations of these potentials. In that sense, this study is complementary, to other analysis done in finite nuclei~\cite{les07,Ang11,fia02} regarding tensor terms. 
Whereas all the microscopic calculations are in quite reasonable agreement at saturation density and in the low-density region~\cite{bal12}, presently only the Brueckner-Hartree-Fock calculations~\cite{bal97} provide us with all partial waves contributions to the EoS in a wide density range. For this reason, BHF results are used to constrain the spin-orbit and tensor terms of effective interactions. It is however implicit that the method drawn in this article would also hold if some other microscopic results were available and used.

The bare nucleon-nucleon interaction contains an important tensor part, which is necessary
to reproduce not only the phase shifts of the nucleon-nucleon scattering, but also the quadrupole
moment of the deuteron. However, apart from some exploratory studies~\cite{sta77,oni78}, these terms have
been omitted in the study of finite nuclei up to recently. In general, the inclusion of tensor terms allows for a better description of the evolution of nuclear shells for stable nuclei as well as exotic ones. Furthermore, it has received a particular attention in the recent years due to its contribution to the shell evolution in atomic nuclei~\cite{ots06}.
Nowadays there are several zero- and finite-range effective interactions containing tensor terms. Among the finite-range ones, let us mention the M3Y type effective interaction by Nakada~\cite{nak03,nak08,nak13}, which includes finite-range spin-orbit and tensor terms, whose parameters are fitted to the Brueckner's G-matrix based on a bare interaction. Much closer to the spirit of an effective interaction, a tensor term was added to a Gogny-type potential to describe the evolution of
nuclear shells for exotic nuclei as well as stable ones~\cite{Ang11,Ang12}. Concerning Skyrme interactions, zero-range tensor terms as originally proposed by Skyrme have been either included perturbatively to existing central ones or with
a complete refit of the parameters~\cite{les07}. For a general discussion see Ref.~\cite{sag14}, where a
systematic study of the zero-range effective tensor interaction combined with a standard Skyrme
interaction has been made.   

The link between zero and finite-range interactions is the object of some debate. In this article we contribute to it, trying in particular to show how a zero-range interaction can constitute a reliable approximation of a finite-range one.
This leads us to consider the next-to-next-to-next to leading order momentum expansion of the Skyrme pseudo-potential~\cite{car08,rai11} which represents an extension of the standard Skyrme model~\cite{sky59,kor13}. 
It is currently referred to as N3LO Skyrme pseudo-potential, a term introduced in Ref.~\cite{car08,rai11} in analogy with expansion techniques employed in chiral effective field theory methods.  
The role of the higher order gradients in N3LO is to mimic the effect of a range with higher and higher accuracy (in agreement with Ref.~\cite{car10} in the context of the Density Matrix Expansion). In this respect, we have obtained the N3LO zero-range limit of the Gogny and M3Y interactions. Thanks to the analytical properties of the infinite nuclear medium, we will show that the parameters of such an extended Skyrme interaction can be related in a simple way to the ones of Gogny or M3Y. By performing a systematic study of the partial wave decomposition, we will also discuss how the main contributions to the EoS come from the partial waves $S,P,D,F$~\cite{Dav16PRC} which are naturally included into the N3LO pseudo-potential. 

The article is organized as follows: after recalling for completeness in Sect.~\ref{sect:interactions} the main characteristics of the chosen finite-range interactions, we present in Sect.~\ref{sect:decomposition} the calculated equations of state for symmetric nuclear matter, and examine their decomposition in terms of $(S,T)$ channels and  partial waves. Explicit formulae are given and the results obtained from several parametrizations are compared with BHF results. In Sect.~\ref{sect:finitezero}, we address the question of gauge invariance of the finite-range spin-orbit interaction term. Then, we determine the zero-range limit of the finite-range interactions and discuss the advantages of the N3LO pseudo-potential. The general conclusions are presented in Sect.~\ref{sect:conclusions}. 
The complete set of equations used in the article are presented in the Appendices.

\section{Effective NN interactions}
\label{sect:interactions}

In this Section, we recall the main characteristics of the effective interactions considered in the present article, namely the finite-range Gogny and M3Y ones, and the N3LO Skyrme interaction. 

\subsection{Gogny interaction}
The effective Gogny interaction was proposed in the seventies aiming at offering a fair description of static properties of spherical as well as deformed nuclei. In its standard form~\cite{dec80}, it is a sum of central, spin-orbit and density-dependent terms. The inclusion of an explicit tensor term has been proposed by the authors of Refs.~\cite{Ang11,Ang12,Gra13}. In the present work we will thus consider such an extended form of the Gogny interaction~\cite{ber16}
%%%
\begin{eqnarray}\label{eq:gog}
v_{G}(\mathbf{r}_1,\mathbf{r}_2)=v^C_{G}+v^{LS}_{G}+v^{DD}_{G}+v^{T}_{G} .
\end{eqnarray}
The subindex $G$ stands for Gogny. 
%%%
The central term consists of a sum of two Gaussian functions
%%%
\begin{equation}\label{gog:cen}
v^C_{G}(\mathbf{r}_1,\mathbf{r}_2) = \sum_{i=1}^2 \left[ W_i+B_iP_{\sigma}-H_iP_{\tau}-M_iP_{\sigma}P_{\tau}  \right] e^{-(r / \mu^{C}_i)^2} ,
\end{equation}
%%%
the choice of the ranges depending on the adopted parametrisation.
Both the spin-orbit and the density dependent terms have the same form as the standard Skyrme interaction~\cite{vau72}
%%%
\begin{eqnarray}\label{gog:so}
v^{LS}_{G}(\mathbf{r}_1,\mathbf{r}_2) &=& iW_0(\sigma_1+\sigma_2)\cdot\left[ \mathbf{k}'\times \delta(\mathbf{r})\mathbf{k}\right] , \\
\label{gog:dd}
v^{DD}_{G}(\mathbf{r}_1,\mathbf{r}_2) &=& t^{(DD)}(1+x^{(DD)} P_{\sigma})\rho^{\alpha}(\mathbf{R})\delta(\mathbf{r}) ,
\end{eqnarray}
%%%
where $\mathbf{k}= i \; (\nabla_1-\nabla_2) /2 $ is the relative momentum operator acting on the right and $\mathbf{k}'$ is acting on the left, and $\mathbf{R}=(\mathbf{r}_1+\mathbf{r}_2)/2$. $ P_{\sigma}, P_{\tau}$ are the spin/isospin exchange operators~\cite{ben03}.
The density-dependent form, with parameter $\alpha=1$, was suggested in Ref.~\cite{vau72} in replacement of the original zero-range three-body term~\cite{sad13}. For that historical reason, the global coefficient in the standard form is usually written as $t_3/6$. 
Afterwards it was shown that a value of $\alpha$ less than 1 is required to get an acceptable SNM compressibility.
In the original Gogny's notation~\cite{dec80} the parameters were denoted $t_0, x_0$, written here as $t^{(DD)}, x^{(DD)}$ to avoid posterior confusions with the Skyrme interaction.  

\noindent Finally, for the tensor term we consider the form recently proposed in Ref.~\cite{Gra13}
%%%
\begin{eqnarray}\label{gog:ten}
v^{T}_{G}(\mathbf{r}_1,\mathbf{r}_2) = (V_{T1}+V_{T2}P_{\tau})e^{-(\mathbf{r} / \mu^T )^2} r^2 S_{12} ,
\end{eqnarray}
%%%
where the tensor operator is defined as
%%%
\begin{eqnarray}
S_{12}=4\left[ 3(\mathbf{s}_1\cdot \hat{\mathbf{r}})(\mathbf{s}_2\cdot \hat{\mathbf{r}})-\mathbf{s}_1\cdot\mathbf{s}_2\right] , 
\end{eqnarray}
%%%
with $\mathbf{s}_{1},\mathbf{s}_{2}$ are the nucleon spin operators. 
The parameters $V_{T1},V_{T2}$ have not been fitted together with the other parameters, but adjusted without any modifications of the central part parameters. In Ref.~\cite{Gra13} for instance, the spin-orbit parameter has been modified together with the tensor term. In that sense, we can say that the tensor has been added perturbatively to the Gogny interaction contrary to the M3Y one for which the tensor parameters enter directly in the fitting procedure.

Several parametrizations have also been fitted~\cite{dec80,gor09b,cha08} aiming at improving the description of finite nuclei energies and sizes, energy levels, pairing properties, nucleon matter, etc. Finally, a new interaction named D2, with a finite-range version of the density dependent term has been presented recently in Ref.~\cite{cha15}. Our results can be easily modified to take into account such a new term.

\subsection{M3Y interaction}
This interaction has been initially devised~\cite{ber77} to study inelastic scattering aiming at relying it to bare NN interactions through their $G$-matrix elements, which were fitted to as sum of Yukawa functions. This effective interaction has been extended recently to nuclear structure studies by Nakada~\cite{nak03,nak08,nak13} with the addition of a density-dependent term to get correct values for saturation density in symmetric nuclear matter. This version will be considered along this article. The M3Y interaction contains central, 
tensor, spin-orbit and density-dependent terms
%%%
\begin{eqnarray}\label{eq:m3y}
v_{N}(\mathbf{r}_1,\mathbf{r}_2)=v^C_{N}+v^{T}_{N}+v^{LS}_{N}+v^{DD}_{N} ,
\end{eqnarray}
where the subindex $N$ stands for Nakada.
%%%
The central part is the sum of three Yukawa functions
%%%
\begin{equation}\label{nak:cen}
v^C_{N}(\mathbf{r}_1,\mathbf{r}_2) = \sum_{i=1}^3 \left[ t_i^{(SE)}P_{SE} +t_i^{(TE)}P_{TE}+t_i^{(SO)}P_{SO}+t_i^{(TO)}P_{TO}  \right] \frac{e^{-r\; \mu^{C}_i}}{r \; \mu^{C}_i} ,
\end{equation}
%%%
where the relative coordinate is $\mathbf{r}=\mathbf{r}_1-\mathbf{r}_2$,  and the  operators $P_{SE}, P_{TE}, \dots$ project onto singlet-even, triplet-even, $\dots$ two-particle states, and are given in terms of the spin and isospin exchange operators $P_{\sigma}$ and $P_{\tau}$ as
%%%
\begin{eqnarray}
 P_{SE}=\frac{1}{4}(1-P_{\sigma})(1+P_{\tau}) \, , &&  P_{TE}=\frac{1}{4}(1+P_{\sigma})(1-P_{\tau})  ,\nonumber \\
 P_{SO}=\frac{1}{4}(1-P_{\sigma})(1-P_{\tau}) \, , &&  P_{TO}=\frac{1}{4}(1+P_{\sigma})(1+P_{\tau})  . \nonumber
\end{eqnarray}
%%%
The ranges $\mu^{C}_{i}$ have been fixed in Ref.~\cite{ana83} by adjusting the interaction M3Y-P0 on the results extracted from Paris-potential.

The tensor and the spin-orbit terms are a sum of two Yukawa functions
%%%
\begin{eqnarray}\label{nak:ten}
v^T_{N}(\mathbf{r}_1,\mathbf{r}_2)&=&\sum_{i=1}^2 \left[ t_i^{(TNE)}P_{TE} +t_i^{(TNO)}P_{TO}  \right] \frac{e^{-r \, \mu^{T}_i}}{r \, \mu^{T}_i} r^2 S_{12} ,\\
v^{LS}_{N}(\mathbf{r}_1,\mathbf{r}_2)&=&\sum_{i=1}^2 \left[ t_i^{(LSE)}P_{TE} +t_i^{(LSO)}P_{TO}  \right] \frac{e^{-r \,\mu^{LS}_i}}{r \, \mu^{LS}_i}\mathbf{L}_{12}\cdot (\mathbf{s}_1+\mathbf{s}_2) .
\end{eqnarray}
%%%
The relative orbital angular momentum operator 
is $\mathbf{L}_{12}=\mathbf{r} \times\mathbf{p}$, where $\mathbf{p}=(\mathbf{p}_1-\mathbf{p}_2)/2$ is the relative momentum.
% It is worth noticing that the ranges of the Yukawa potential for tensor and spin-orbit $\mu_n^{T},\mu_n^{LS}$ are not the same as for the central term~\cite{nak03}. 

Finally, a zero-range density-dependent term is added 
%%%
\begin{eqnarray}
v^{DD}_{N}(\mathbf{r}_1,\mathbf{r}_2) = t^{(DD)}(1+x^{(DD)}P_{\sigma}) \, \rho^{\alpha}(\mathbf{r}_1) \, \delta(\mathbf{r}) .
\end{eqnarray}
%%%

The most recent  M3Y parametrizations~\cite{nak08} use a combination of two such terms with different density powers for the single-even and triplet even contribution \cite{nak13} .  
It is worth noticing that a density dependent term has also been added recently to the spin-orbit term~\cite{Nak15} to mimic the effect of three-nucleon interactions. We will not consider such a term in the present discussion.

\subsection{N3LO}

We now recall the structure of the N3LO Skyrme pseudo-potential ~\cite{car08,rai11,dav14c}. It is a generalisation of the standard Skyrme interaction, corresponding to the expansion of the momentum space matrix elements of a generic interaction in powers of $\mathbf{k}, \mathbf{k}'$  up to the sixth order, imposing Galilean and local gauge invariance~\cite{dob95}. It is written as 
%%%
\begin{eqnarray}
V_{\rm N3LO} =V_{\rm N3LO}^{C}+V_{\rm N3LO}^{LS}+V_{\rm N3LO}^{DD}+V_{\rm N3LO}^{T} .
\end{eqnarray}
%%%
The central term reads
%%%
\begin{eqnarray} \label{eq:N3LO:c}
V_{\rm N3LO}^{C} &=& t_0 (1+x_0 P_{\sigma}) + \frac{1}{2} t_1 (1+x_1 P_{\sigma}) ({\bf k}^2 + {\bf k'}^2) + t_2 (1+x_2 P_{\sigma}) ({\bf k} \cdot {\bf k'})  \nnn
            & & + \frac{1}{4} t_1^{(4)} (1+x_1^{(4)} P_{\sigma}) \left[({\bf k}^2 + {\bf k'}^2)^2 + 4 ({\bf k'} \cdot {\bf k})^2\right] \nnn
            & &+ t_2^{(4)} (1+x_2^{(4)} P_{\sigma}) ({\bf k'} \cdot {\bf k}) ({\bf k}^2 + {\bf k'}^2) \nnn
& & + \frac{t_1^{(6)}}{2}(1+x_1^{(6)}P_\sigma) ({\bf k}^2 + {\bf k'}^2)\left[({\bf k}^2 + {\bf k'}^2)^2 + 12 ({\bf k'} \cdot {\bf k})^2\right] \nnn
&& + t_2^{(6)}(1+x_2^{(6)}P_\sigma)
({\bf k'} \cdot {\bf k}) \left[3({\bf k} ^2 + {\bf k'}^2)^2 +4({\bf k'} \cdot {\bf k})^2\right] .
\end{eqnarray}
%%%
In these expressions, a $\delta({\bf r}_1-{\bf r}_2)$ function is to be understood, but has been omitted for the sake of clarity. See Ref.~\cite{ben03} for details on the adopted notation. The spin-orbit and density-dependent terms have exactly the same structure of the standard Skyrme interaction~\cite{cha97}.

The tensor term reads 
%%%
\begin{eqnarray} \label{eq:N3LO:t}
V_{\rm N3LO}^{T} &=& \frac{1}{2} t_e T_e({\bf k'},{\bf k})  + \frac{1}{2} t_o T_o({\bf k'},{\bf k})  \nnn
& & + t_e^{(4)} \left[  ({\bf k}^2+{\bf k'}^2) T_e({\bf k'},{\bf k}) +  2 ({\bf k'} \cdot {\bf k}) T_o({\bf k'},{\bf k})  \right] \nnn
& & + t_o^{(4)} \left[ ({\bf k}^2+{\bf k'}^2) T_o({\bf k'},{\bf k})  
+  2 ({\bf k'} \cdot {\bf k}) T_e({\bf k'},{\bf k}) \right] \nnn
& & + t_e^{(6)} \left[ \left(\frac{1}{4}({\bf k} ^2 + {\bf k'}^2)^2 +({\bf k'} \cdot {\bf k})^2\right) T_e({\bf k'},{\bf k})
+ ({\bf k} ^2 + {\bf k'}^2) ({\bf k'} \cdot {\bf k})  T_o({\bf k'},{\bf k}) \right] \nnn
& & +t_o^{(6)} \left[\left(\frac{1}{4}({\bf k} ^2 + {\bf k'}^2)^2 +({\bf k'} \cdot {\bf k})^2\right) T_o({\bf k'},{\bf k})
+ ({\bf k} ^2 + {\bf k'}^2) ({\bf k'} \cdot {\bf k})  T_e({\bf k'},{\bf k}) \right] , 
\end{eqnarray}
%%%
where $T_e$ and $T_o$, respectively even and odd under parity transformations, are defined as
%%%
\begin{eqnarray}
T_e({\bf k'},{\bf k}) &=& 3 (\vec \sigma_1 \cdot {\bf k'}) (\vec \sigma_2 \cdot {\bf k'}) 
+ 3 (\vec \sigma_1 \cdot {\bf k}) (\vec \sigma_2 \cdot {\bf k}) 
- ({\bf k'}^2 + {\bf k}^2) (\vec \sigma_1 \cdot \vec \sigma_2), \\
T_o({\bf k'},{\bf k}) &=& 3 (\vec \sigma_1 \cdot {\bf k'}) (\vec \sigma_2 \cdot {\bf k}) 
+ 3 (\vec \sigma_1 \cdot {\bf k}) (\vec \sigma_2 \cdot {\bf k'}) 
- 2 ({\bf k'} \cdot {\bf k}) (\vec \sigma_1 \cdot \vec \sigma_2).
\end{eqnarray} 
%%%
Up to now there is no parametrisation of the N3LO interaction which incorporates the constraints of finite nuclei: the only available parametrizations have been derived from properties of the infinite medium~\cite{Dav15,dav14c,Dav15AA,Dav16}, such as the Landau parameters of neutron matter~\cite{hol13}.

\section{Equation of state and its decompositions}
\label{sect:decomposition} 
 
The energy per particle of an infinite medium is calculated here within the Hartree-Fock (HF) approximation as
%%%
\begin{eqnarray}\label{eq:hf}
\frac{E}{A}=\frac{3}{5}\frac{\hbar^2 k_F^2}{2m}+ {\cal V} , 
\end{eqnarray}
%%%
where $k_F$ is the Fermi momentum, related to the density $\rho$ by $k_F=(3 \pi^2 \rho/2)^{1/3}$ for symmetric nuclear matter and $k_F=(3 \pi^2 \rho)^{1/3}$ for pure neutron matter. 
The potential energy per particle is given by
\begin{equation}
\label{eq:potener}
{\cal V} = \frac{1}{2 A}\sum_{ij}\langle ij| V|ij - ji\rangle .
\end{equation}
In the above equation, the antisymmetric matrix element of the potential interaction has to be calculated on a plane wave basis; indices $i, j$ span the occupied states in the Fermi sphere. The calculation is standard, and the resulting expressions are given in \ref{app:EOS}. Detailed expressions for Gogny interaction can also be found in  Ref.~\cite{sel14}.

%%%%%%%%%%%%%%%%%%%%%%%%%%%%%%%%%%%%%%%%%%%%%%%%%%%%%%%
\begin{figure*}[h] 
   \centering
   \includegraphics[angle=-90,width=0.48\textwidth]{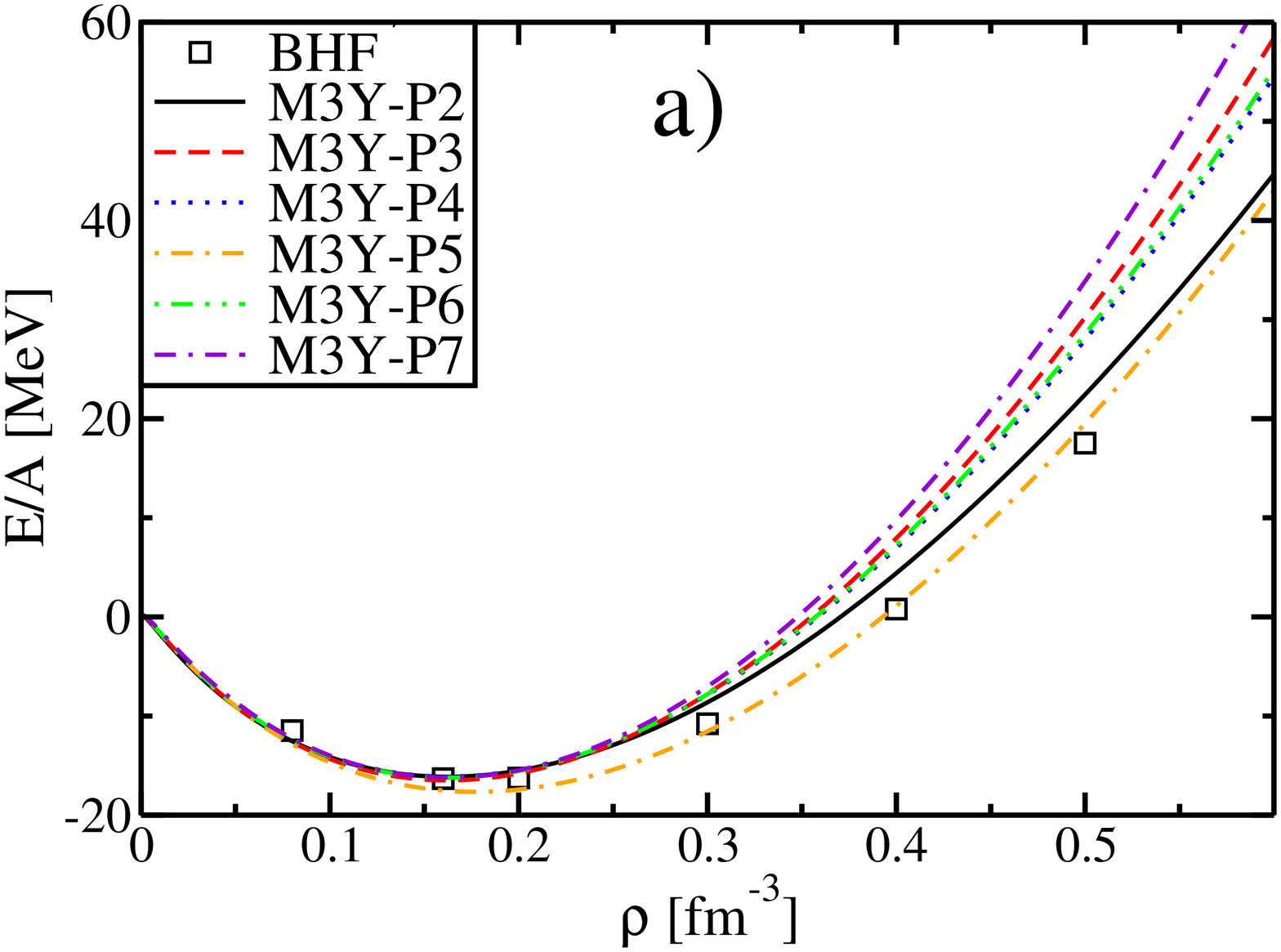}
   \includegraphics[angle=-90,width=0.48\textwidth]{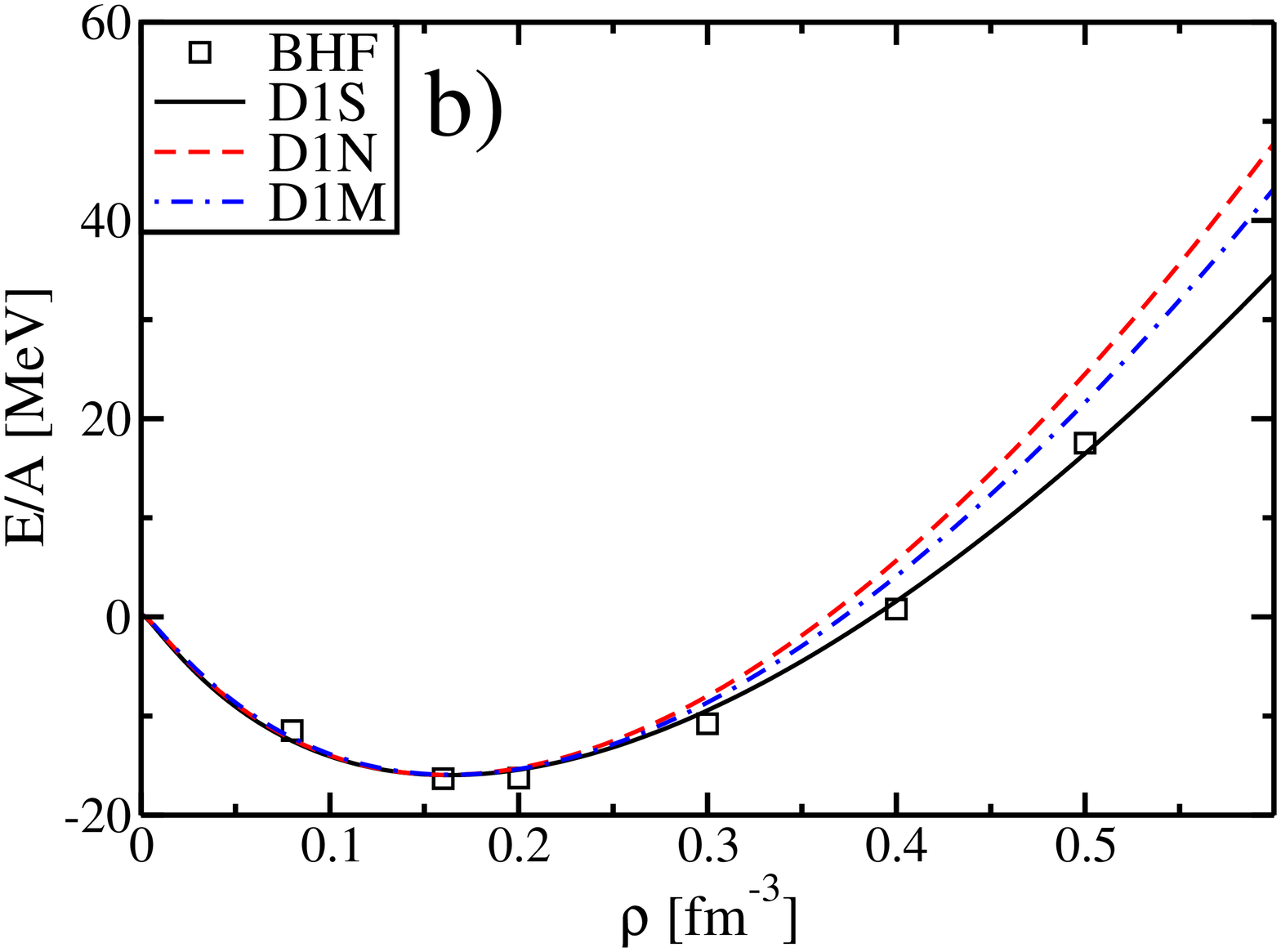}
   \caption{(Colors online) Equations of state of SNM  for the M3Y (panel a)  and Gogny (panel b) parametrizations. The lines represent the result obtained with the effective interactions discussed in this article while the empty squares represent the EoS obtained with BHF methods \cite{bal97}.}
   \label{fig:EoS}
\end{figure*}
%%%%%%%%%%%%%%%%%%%%%%%%%%%%%%%%%%%%%%%%%%%%%%%%%%%%%%%

The energy per particle depends only on the central and density-dependent parts of the effective interaction: neither the spin-orbit nor the tensor terms contribute to the EoS of a saturated system. Nevertheless, it is possible to separate explicitly their contributions by means of a partial wave decomposition of the total EoS, as it will be shown in the next subsection.

It is worth remembering that there is an important difference between the simple HF based on the effective interactions and the more complex BHF methods: the calculation of the $G$-matrix involves self-energies and intermediate states energies thus including contributions of non-central terms~\cite{vid11}. These effects can not be mimicked by an effective interaction in a simple way at HF level. For such a reason, we will not try to separate them and we will concentrate only on their total effect on the partial wave decomposition.
In Fig.~\ref{fig:EoS}, we show the results obtained with several parametrizations for the symmetric nuclear matter EoS, 
together with the \emph{ab-initio} results given in Ref.~\cite{bal97} and based on the BHF calculations.
As expected, all these parametrizations agree for densities below and around the saturation value $\rho_0$, which is the relevant density interval for the description of finite nuclei. Sizeable differences appear only at about two times the saturation density and beyond. We see that both families of interactions give a fair description of the BHF results, but the Gogny interaction gives a better description at high densities. 
The only available existing constraints on the EoS in the density range $\rho/\rho_0\in[1.5,4.5]$ come from  heavy ion collisions experiments~\cite{dan02,lyn09}. The BHF results agree with these constraints up to these densities and thus we can consider them reliable up to $\rho\approx0.4~$fm$^{-3}$~\cite{bur07}. 
We refer the reader to Ref.~\cite{bal12} for a detailed discussion on the comparison between different {\it ab-initio} methods to calculate the EoS in SNM.
The BHF have been extended to even larger values of the density to be used as a guidance for the production of EoS for astrophysical purposes~\cite{bal97,zho04} and thus used as a reference for constraining effective interactions~\cite{Dav15AA}. The reliability of such extrapolations is still under debate and we stress that the results presented here are not related to the particular choice of the BHF results.
The only available constraint at such high density is the possibility of sustaining a massive neutron star~\cite{dem10}.

\begin{figure}[h]
   \centering
   \includegraphics[angle=-90,width=0.48\textwidth]{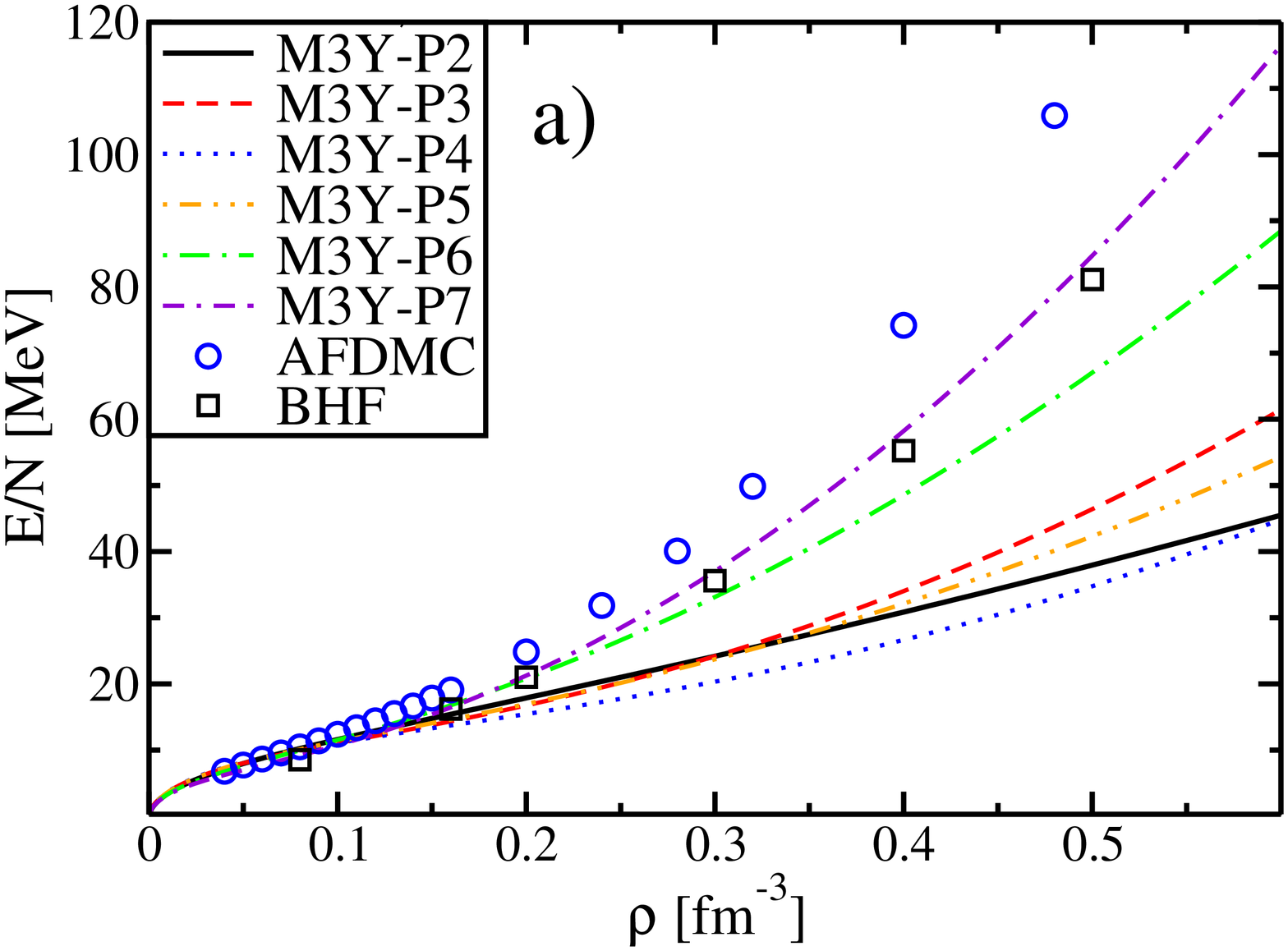}
   \includegraphics[angle=-90,width=0.48\textwidth]{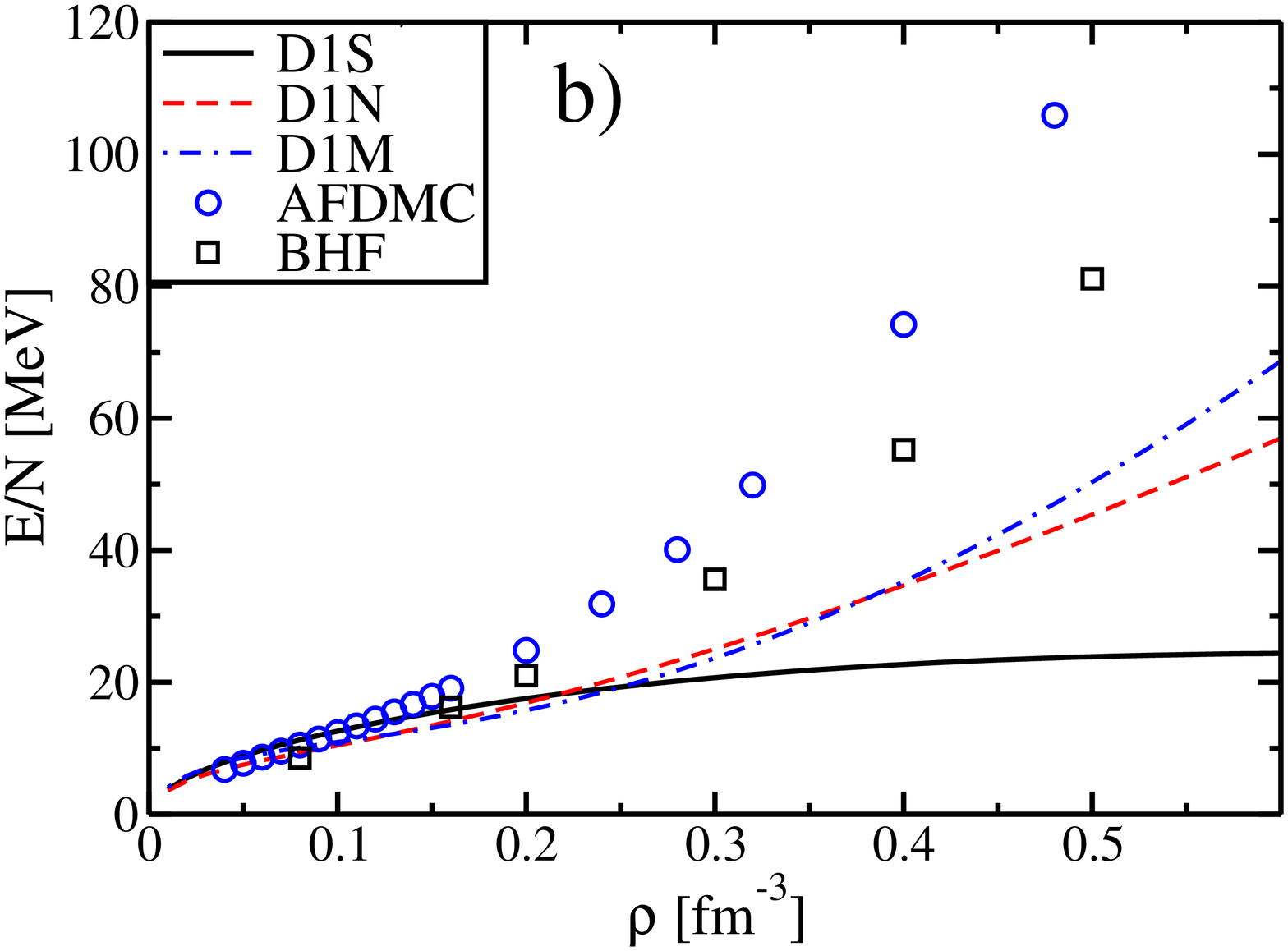}
   \caption{(Colors online) Same as Fig.~\ref{fig:EoS} but for PNM.}
    \label{fig:PNM}
\end{figure}

The EoS in SNM is not sufficient to characterise the properties of an effective interaction. Thus, it is also fundamental to consider the important case of extreme isospin asymmetry, $i.e.$ pure neutron matter (PNM), which is a useful constraint for the isovector properties of the interactions~\cite{cha97}.  In Fig.~\ref{fig:PNM} we present the PNM results for the M3Y interactions (panel a) and the  Gogny ones (panel b). We clearly see that beyond 0.2 fm$^{-3}$ there is a large variation in the extrapolation at high density values. We notice that M3Y-P7 follow quite closely the BHF results, since these data enter explicitly in its fitting protocol. Together with BHF calculations, we have also added recent Quantum Monte Carlo simulations based on Argonne AV8' plus UIX interactions~\cite{gan14,car15}. Both calculations agree up to $\rho\approx 0.25$ fm$^{-3}$ but present noticeable differences beyond.% {\bf JN: Je ne comprends pas cette phrase} However, since there is essentially no constraint on the EoS of PNM, the different extrapolations based on different {\it ab initio} methods may give very different results.

%%%%%%%%%%%%%%%%%%%%%%%%%%%%%%%%%%%%%%%%%%%%%%%%%%%%%%%
%%%%%%%%%%%%%%%%%%%%%%%%%%%%%%%%%%%%%%%%%%%%%%%%%%%%%%%

Informations on PNM in a density range of $\rho\approx$~0.1 fm$^{-3}$ can be extracted by comparing the results of several {\it ab-initio} calculations~\cite{dut12} and by comparing the behaviour of dilute Fermi gases~\cite{car03}. See for example the discussion in Ref.~\cite{gor10}.
It is worth remembering that a too soft EoS in PNM, as in the case of Gogny D1S, induces an artificial drift in nuclear binding energies when compared with experimental available masses of neutron rich nuclei. To correct such a behaviour the parametrisation D1N was created~\cite{cha07,cha08}.

\subsection{Decomposition in spin-isospin $(S,T)$ channels}
\label{sect:STchannels} 
Using  the global equation of state, one can only constrain a particular combination of the interaction parameters. If one wants to go one step further and to take into account more infinite matter constraints, the EoS has to be decomposed by using the coupled spin-isospin $(S,T)$ basis as
%%%
\begin{equation}
{\cal V} = \sum_{ST} {\cal{V}^{(ST)}} ,
\end{equation}
%%%
where $\mathcal{V}^{(ST)}$ is the potential energy per particle projected onto the different channels. It should be noticed that since neither the tensor nor the spin-orbit interactions contribute to the $(S,T)$ channels, this decomposition allows us to test and constrain only the central part of the interactions. The analytical results are presented in \ref{app:ST}. 

%%%%%%%%%%%%%%%%%%%%%%%%%%%%%%%%%%%%%%%%%%%%%%%%%%%%%%%
\begin{figure}[H]
   \centering
     \includegraphics[angle=-90,width=0.48\textwidth]{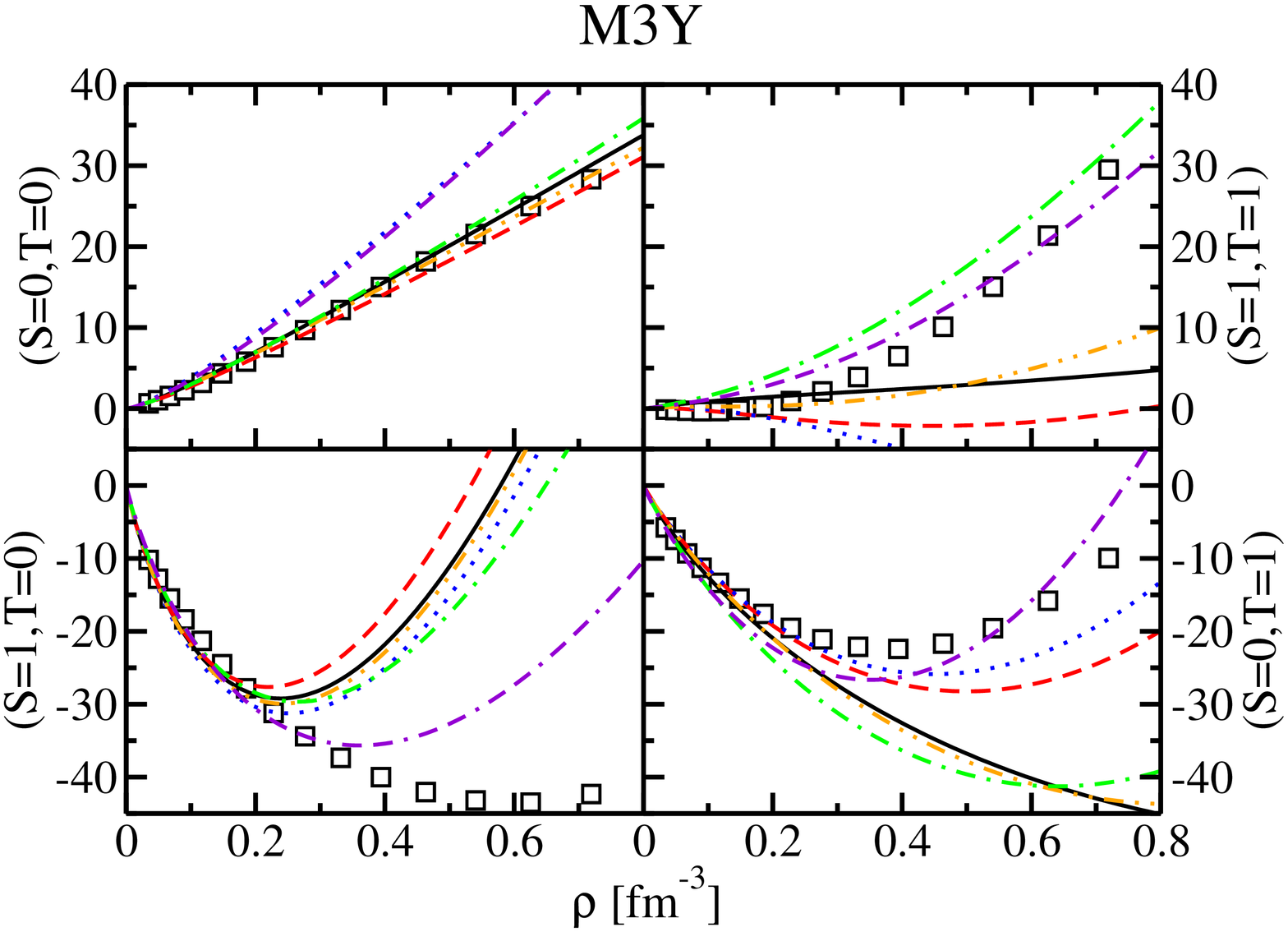}
   \includegraphics[angle=-90,width=0.48\textwidth]{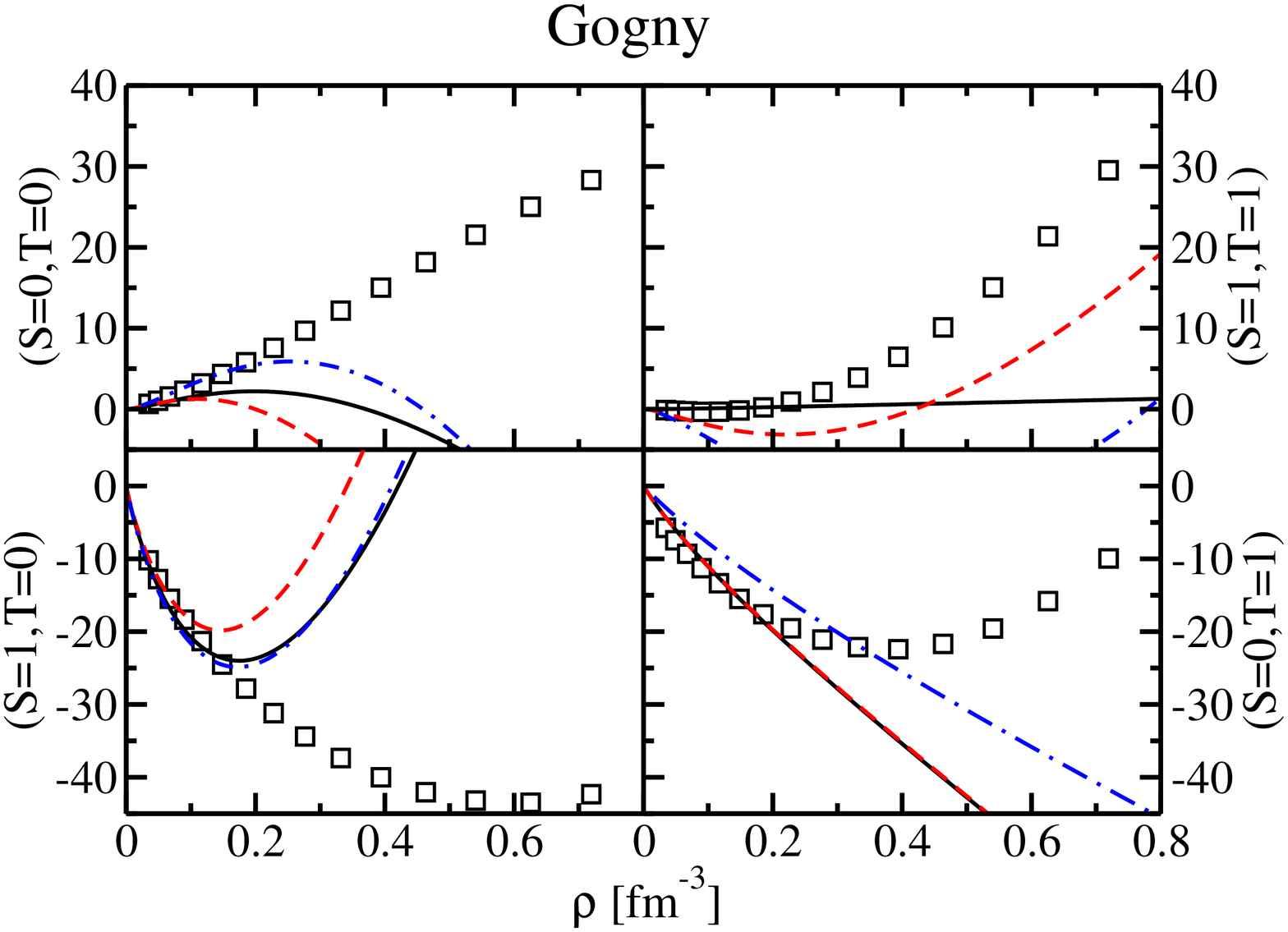}
   \caption{(Colors online) Decomposition of the EoS in SNM for the different $(S,T)$ channels expressed in MeV. On the left panel we present the result for the M3Y interactions and on the right for Gogny ones. The notation is the same as in Fig.\ref{fig:EoS}.}
   \label{fig:STchannel}
\end{figure}
%%%%%%%%%%%%%%%%%%%%%%%%%%%%%%%%%%%%%%%%%%%%%%%%%%%%%%%

In Fig.~\ref{fig:STchannel}, we compare the $(S,T)$-channels for various M3Y and Gogny interactions together with the BHF results. We immediately see that the M3Y interactions are globally in (very) good agreement with the BHF results in the sense that the different parametrizations give the correct behaviour at high densities in all channels and that they all coincide up to saturation density with the BHF results. This means that M3Y interactions contain enough degrees of freedom to satisfy infinite matter and nuclei constraints simultaneously. The same is not true for the Gogny interactions for which we can see that only the original D1S interaction follows the BHF results up to saturation density, while D1N and D1M mainly disagree in the $(S=T=1)$ channel. It is well known that it is nevertheless much better than the standard Skyrme interaction which is not able to give the correct sign in some channels~\cite{les06}.
We recall that the ($S,T$) channels enter explicitly into the fitting procedure of the Gogny interaction~\cite{cha07}, thus we could argue that  two Gaussians (and thus two ranges) are not sufficient. 
 
\begin{figure}[H]
   \centering
   \includegraphics[angle=-90,width=0.6\textwidth]{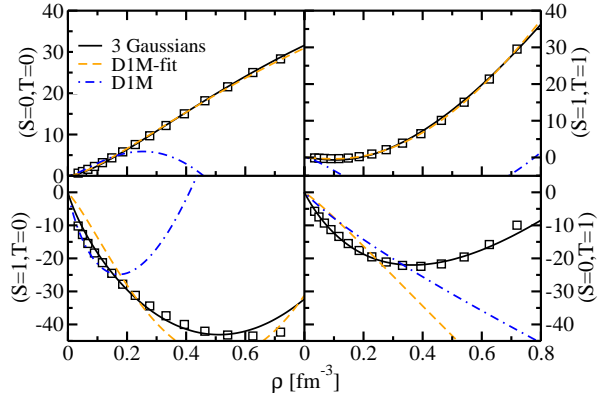}
   \caption{(Colors online) $(S,T)$ channels of the EoS, expressed in MeV, obtained with the Gogny interaction when a third Gaussian is added. The symbols are taken from Ref.~\cite{bal97}.}
   \label{fig:fitG}
\end{figure}

Therefore, we have tested the possibility of improving the reproduction of the $(S,T)$ channels by adding a third Gaussian to the Gogny interaction. We have fixed three different ranges  ($\mu^{C}_1 = 0.25$ fm, $\mu^{C}_2 = 0.8$ fm, $\mu^{C}_3 = 1.2$ fm) and we have then adjusted the parameters on infinite matter properties using the same fitting protocol illustrated in Ref.~\cite{Dav15}. The result is shown in Fig~\ref{fig:fitG}. The very good agreement with the BHF data up to several times the saturation density indicates that the addition of such an extra Gaussian could make the Gogny interaction more suitable for astrophysical calculations~\cite{Dav15AA}.
On the same figure, we also show the original results of D1M interaction and a new fitted version (D1M-fit) where we keep fixed the ranges of the two Gaussian and we re-fit the ($S,T$) channels as in Ref.~\cite{Dav15}.
We notice that even by fitting D1M on $(S,T)$ channels only, thus releasing the finite-nuclei constraints, we still observe that only the $(S=T=0)$ and $(S=T=1)$ channel can be well reproduced.

%%%%%%%%%%%%%%%%%%%%%%%%%%%%%%%%%%%%%%%%%%%%%%%%%%%%%%%
%%%%%%%%%%%%%%%%%%%%%%%%%%%%%%%%%%%%%%%%%%%%%%%%%%%%%%%

\begin{figure}[H]
   \centering
   \includegraphics[angle=-90,width=0.55\textwidth]{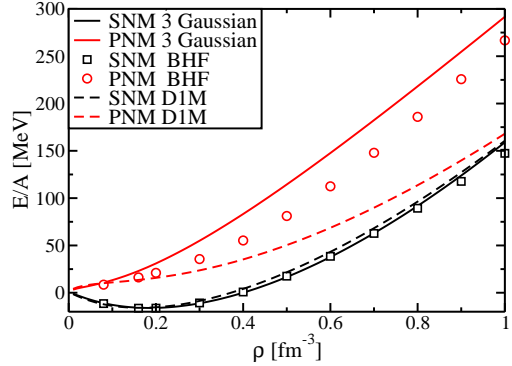}
   \caption{(Colors online) EoS in SNM and PNM obtained with the Gogny interaction when a third Gaussian is added. The symbols are from Ref.~\cite{bal97}.}
   \label{fig:fitGeos}
\end{figure}

However, this improvement is only moderately reflected in the EoS for PNM, as shown in Fig.~\ref{fig:fitGeos}. As compared with BHF results, the softer D1M EoS becomes slightly stiffer when the previous third Gaussian is added. In fact, as already discussed in Ref.~\cite{Dav15AA}, to reproduce with more accuracy the BHF EoS in PNM this should be included in the fit explictly, but the results are very encouraging already at this stage. Moreover, before coming to the more stringent conclusion on the necessity of a third Gaussian, a global fit which incorporates the nuclei is clearly needed.

%%%%%%%%%%%%%%%%%%%%%%%%%%%%%%%%%%%%%%%%%%%%%%%%%%%%%%%
%%%%%%%%%%%%%%%%%%%%%%%%%%%%%%%%%%%%%%%%%%%%%%%%%%%%%%%

%Finally we observe that the M3Y interaction using three ranges Yukawa gives a good description of both ($S,T$) channels and finite nuclei.
%{\bf JN: Conclusion: There is room for improving the central parts, but one should check finite nuclei at the same time. In principle, the same applies for Nakada.}

\subsection{Decomposition in  partial waves}
 \label{sect:partialwaves}
 
To investigate the contribution of spin-orbit and tensor to the EoS, we have to make one step further in the decomposition of the EoS, that is doing a complete partial wave decomposition. This has already been done in Ref.~\cite{dav14k}, but for the N3LO Skyrme interaction~\cite{car08,rai11,dav14c}. Technically speaking, we have to project over a basis with total angular momentum $\vec{J}=\vec{L}+\vec{S}$, with $\vec{L}$ being the total orbital angular momentum and $\vec{S}$ the total spin. All the analytical results can be found in \ref{app:aux}. The global EoS is then obtained by summing up all partial waves contributions as
%%%
\begin{eqnarray}\label{eq:jls:truncated}
{\cal V} = \sum_{JLS}\mathcal{V}(^{2S+1}L_J) ,
\end{eqnarray}
%%%
with the usual spectroscopic notation $^{2S+1}L_J$ for a partial wave with quantum numbers $(S,L,J)$.

\begin{figure}[h] %\label{esura:gog:jlst}
 \centering
      \includegraphics[angle=-90,width=0.5\textwidth]{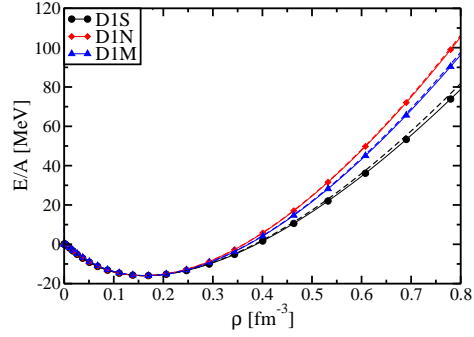}
      \caption{(Colors online)  Comparison between the EoS obtained using full HF calculations for different Gogny interactions, Eq.~\ref{eq:hf}, (solid lines) and the Eq.~\ref{eq:jls:truncated} truncated (dashed-lines)  at $L=3$. }
      \label{Eos:convergence}
\end{figure}

Although a finite-range interaction contributes to all partial waves, a careful inspection of the BHF results~\cite{bal97} shows that the contributions of partial waves with high $L$ ($i.e.$ with $L>3$) to the EoS become less and less important~\cite{vid11}. We thus have neglected partial waves beyond $F$ waves. To support quantitatively this assumption, we compare in Fig.~\ref{Eos:convergence} the total EoS in SNM as given by the complete Gogny interaction, Eq.~(\ref{eq:jls:truncated}), and by truncating it
%HF calculations (Eq.~\ref{eq:hf}) and the one using Eq.~\ref{eq:jls:truncated} with a truncation 
at $L=3$. We clearly observe that these first partial waves are sufficient to reproduce with very good accuracy the total EoS, thus showing that the contribution of higher order partial waves is essentially negligible.

%%%%%%%%%%%%%%%%%%%%%%%%%%%%%%%%%%%%%%%%%%%%%%%%%%%%%%
%%%%%%%%%%%%%%%%%%%%%%%%%%%%%%%%%%%%%%%%%%%%%%%%%%%%%%%

\begin{figure}[H] %\label{esura:gog:jlst}
 \centering
      \includegraphics[angle=-90,width=0.5\textwidth]{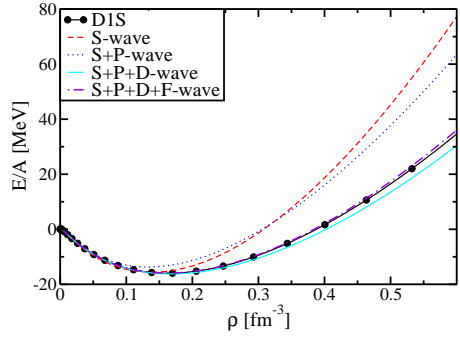}
      \caption{(Colors online) EoS in SNM for the D1S interaction (solid line with dots) and sum of different partial waves.}
      \label{Eos:convergence2}
\end{figure}

In order to have an insight of the relative importance of each partial wave, we show in Fig.~\ref{Eos:convergence2} 
their cumulated contribution to the EoS for the D1S Gogny interaction. Higher values of $L$ are necessary to describe the EoS at large values of density. Fig.~\ref{Eos:convergence2} shows that including only the first three partial waves ($S, P, D$), the exact EoS is reproduced within a few percents up to about 4 times the saturation density.

%%%%%%%%%%%%%%%%%%%%%%%%%%%%%%%%%%%%%%%%%%%%%%%%%%%%%%
%%%%%%%%%%%%%%%%%%%%%%%%%%%%%%%%%%%%%%%%%%%%%%%%%%%%%%%

%%%%%%%%%%%%%%%%%%%%%%%%%%%%%%%%%%%%%%%%%%%%%%%%%%%%%%
\begin{figure}[h] 
 \centering
      \includegraphics[angle=-90,width=0.45\textwidth]{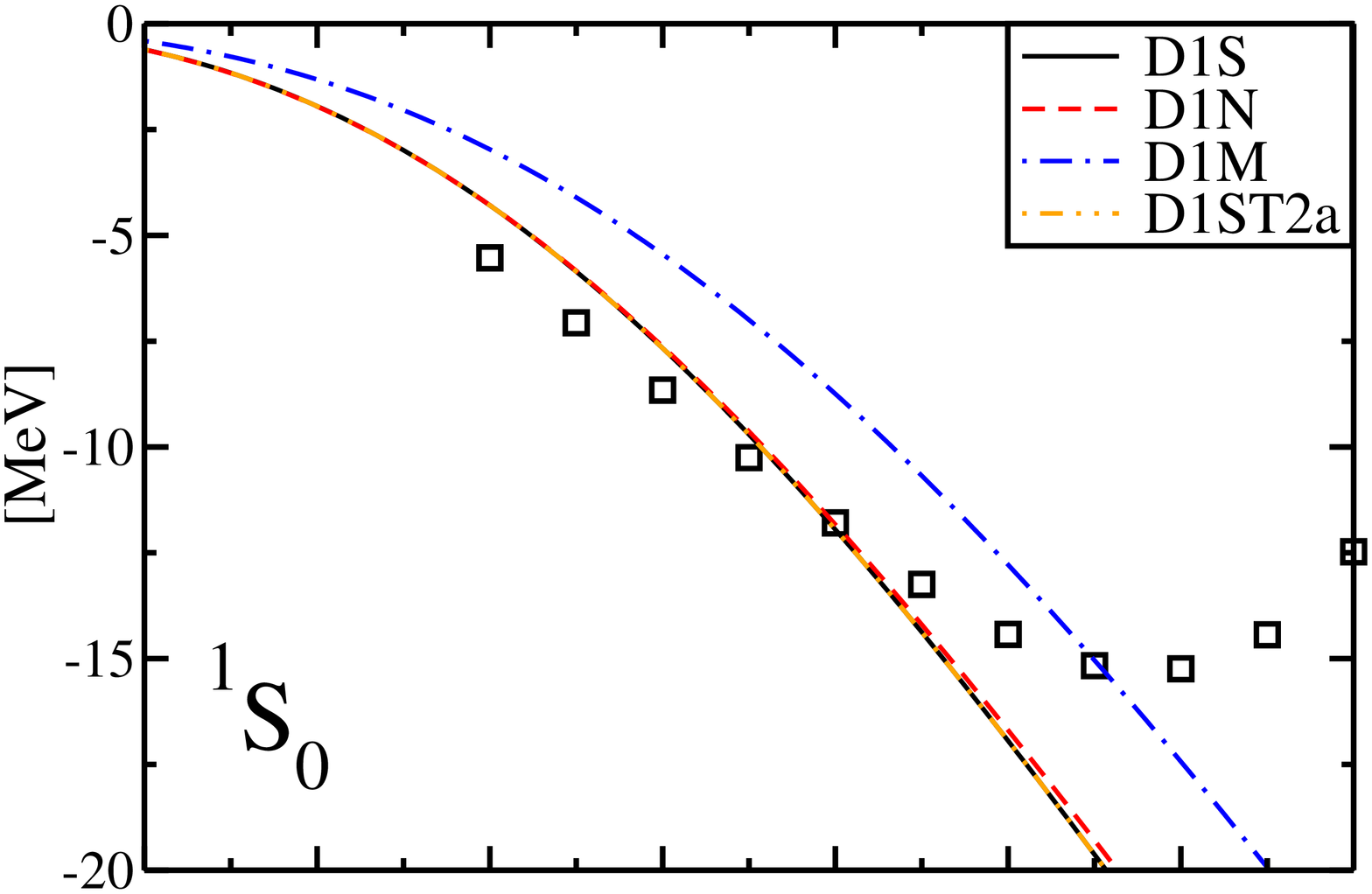}
      \hspace{-15.50mm}
            \includegraphics[angle=-90,width=0.45\textwidth]{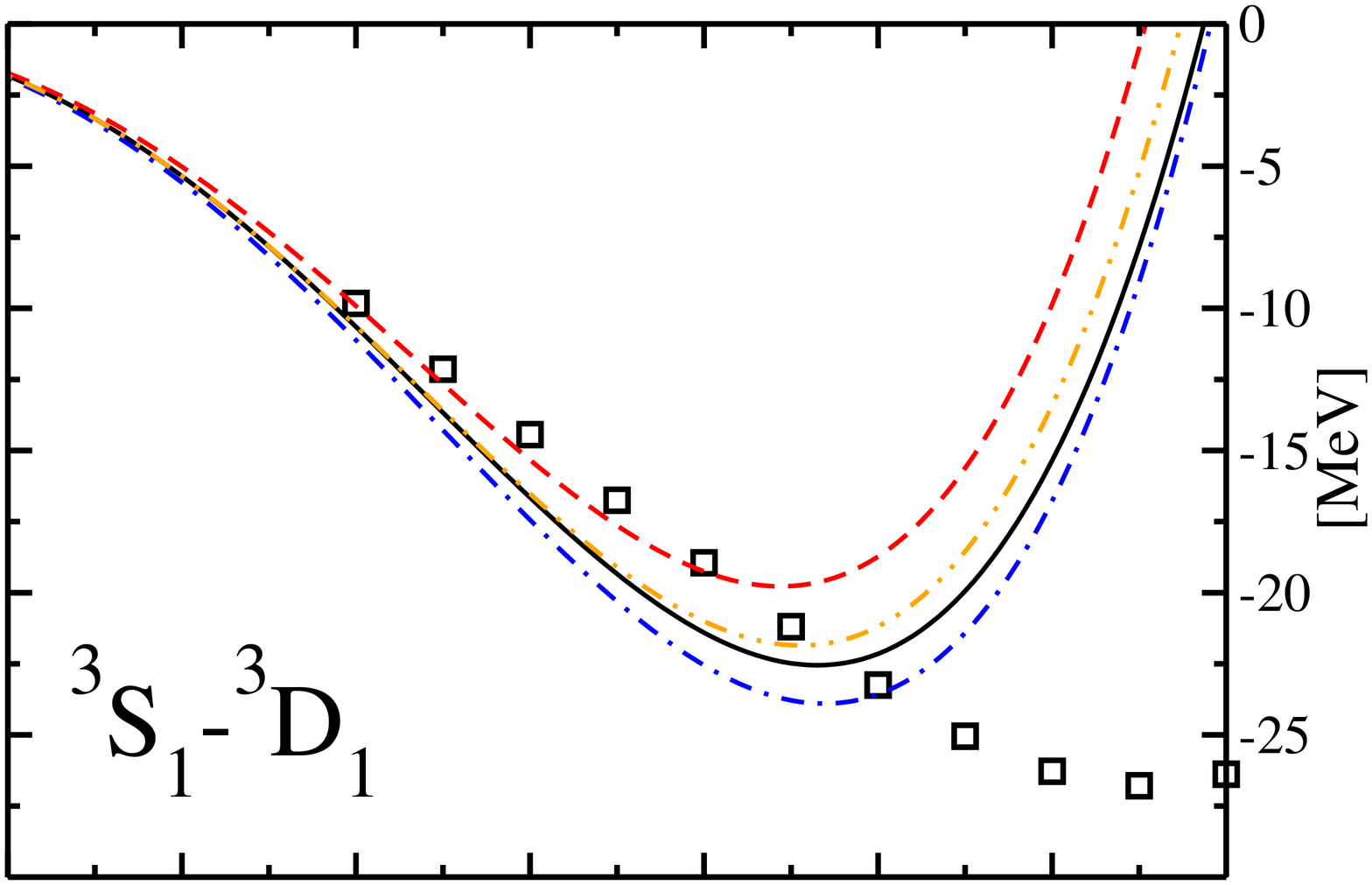}\\
             \vspace{-14.5mm}
                  \includegraphics[angle=-90,width=0.45\textwidth]{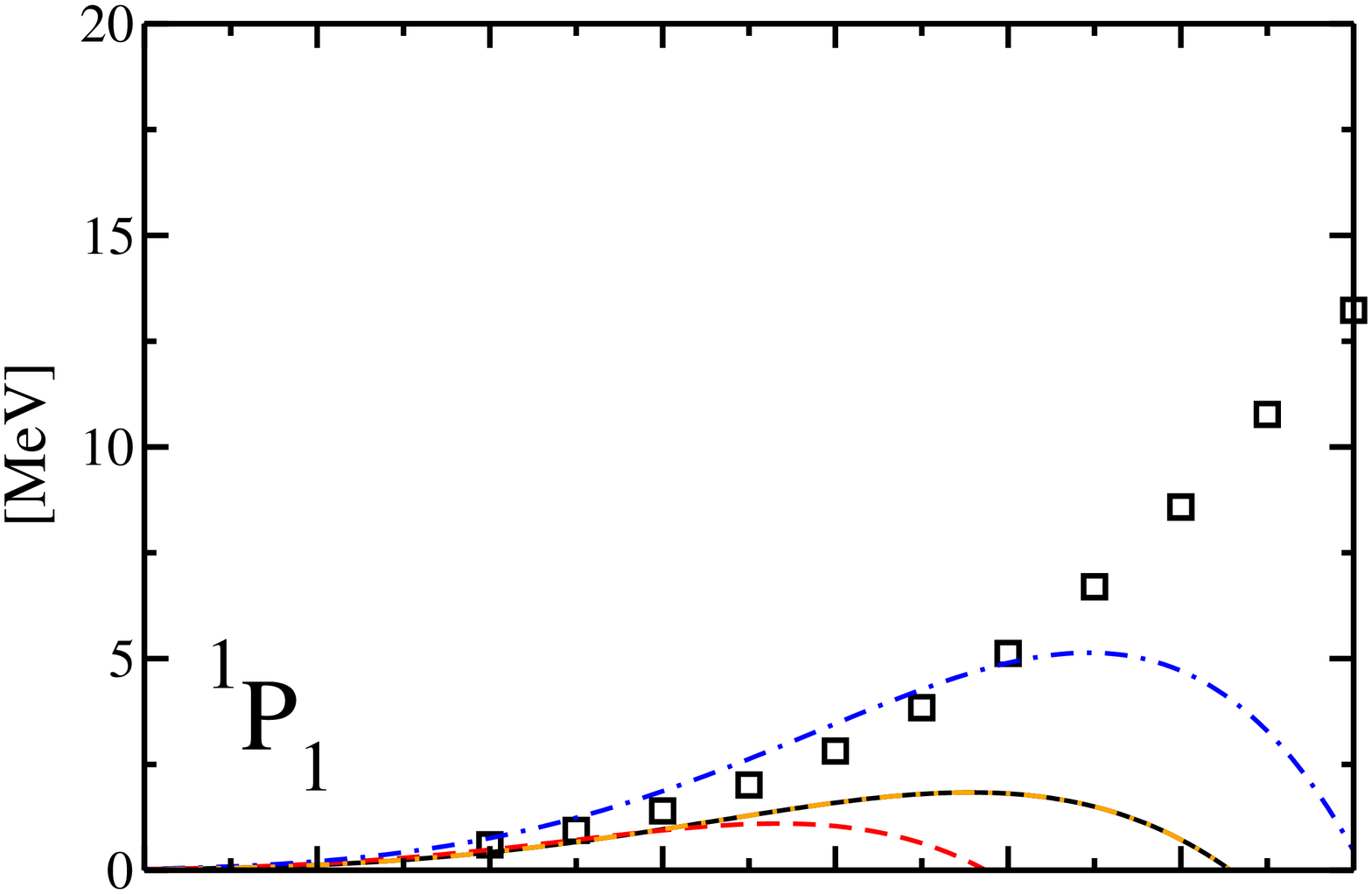}
                     \hspace{-15.50mm}
                  \includegraphics[angle=-90,width=0.45\textwidth]{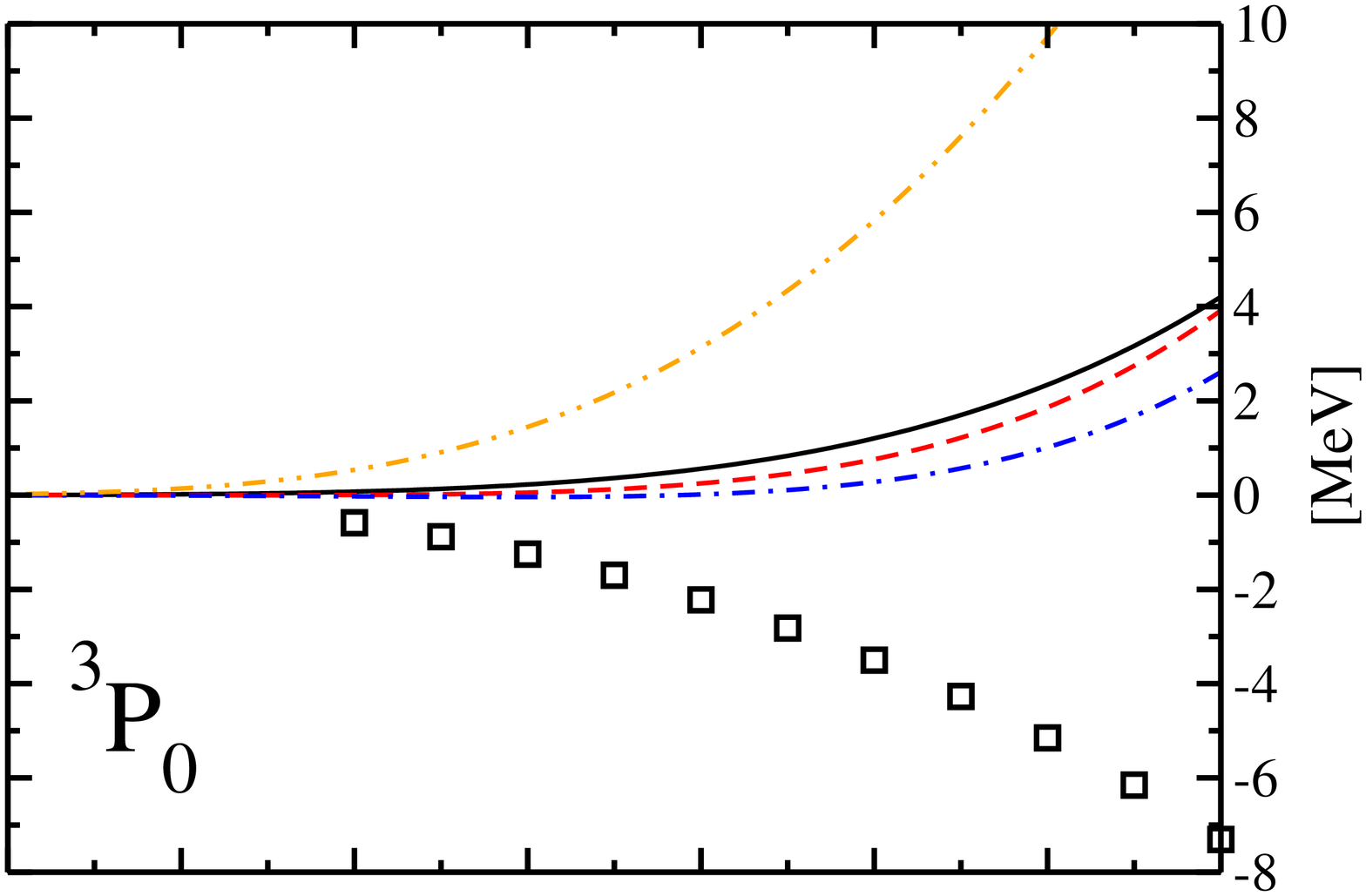}\\
             \vspace{-14.5mm}
                  \includegraphics[angle=-90,width=0.45\textwidth]{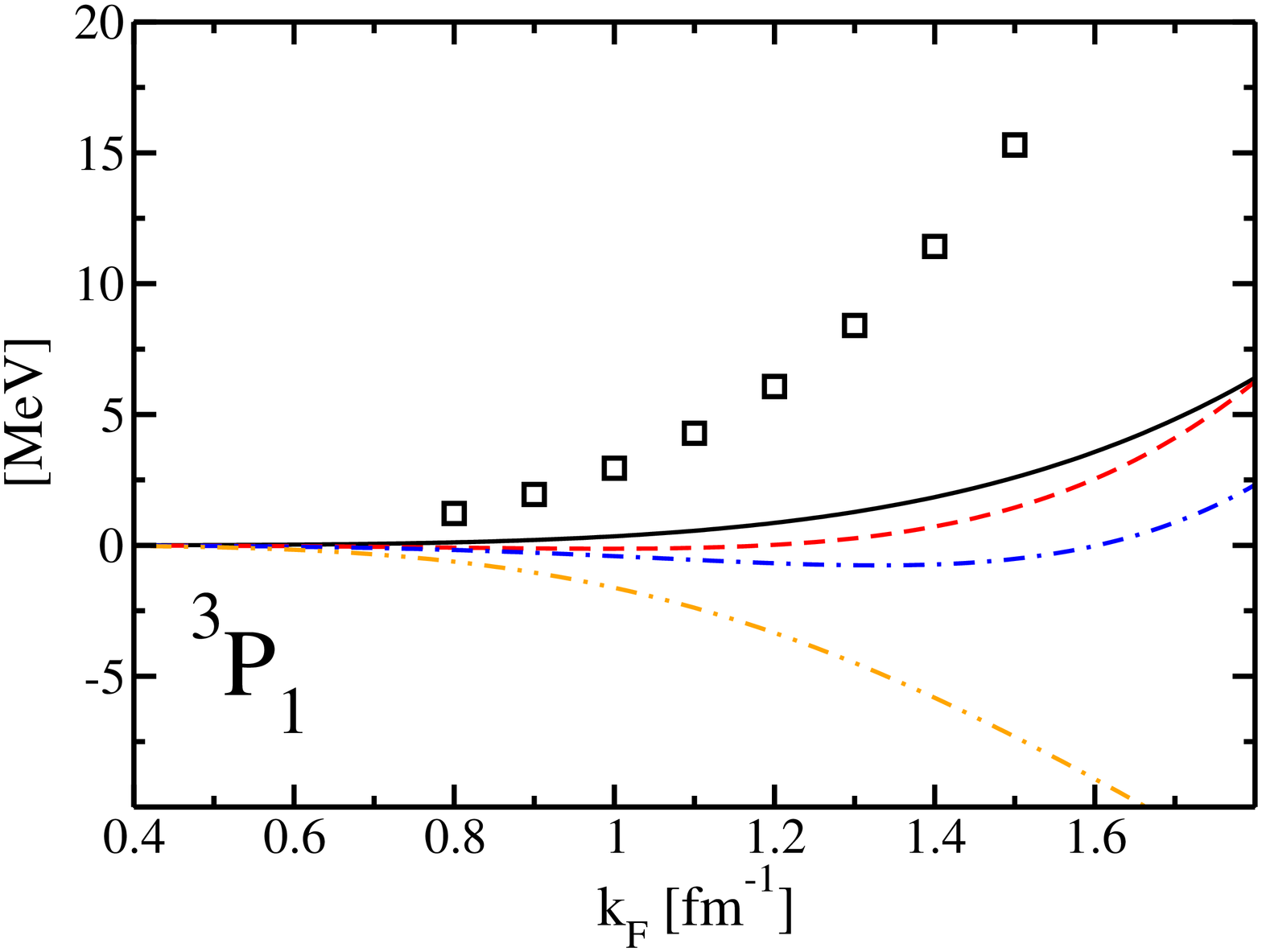}
                    \hspace{-15.50mm}
                  \includegraphics[angle=-90,width=0.45\textwidth]{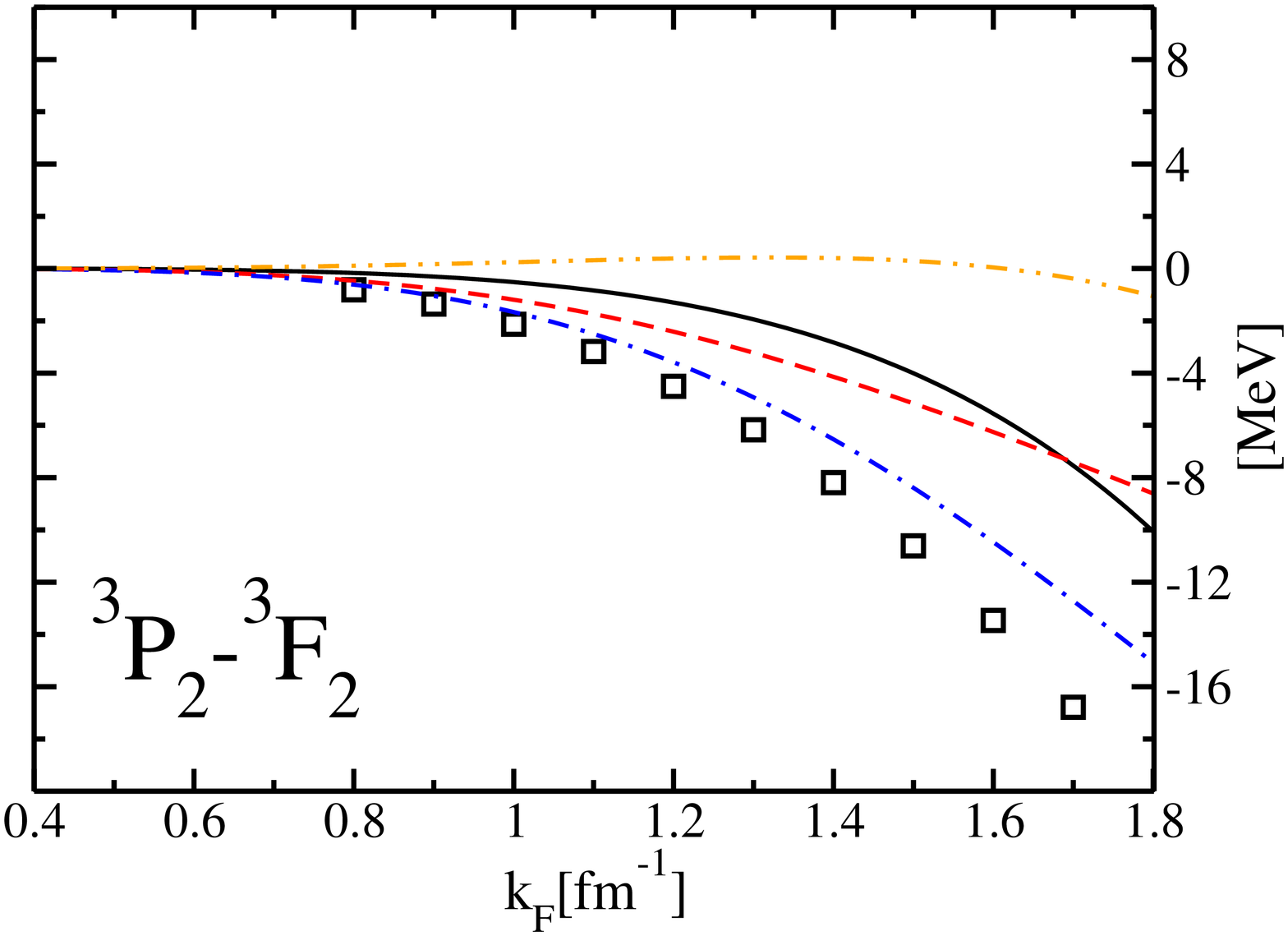}

   \caption{(Colors online) Partial wave decomposition of EoS in SNM for the different Gogny interactions (lines) and the BHF calculations (symbols)~\cite{bal97}.}
   \label{JLST:gog}
\end{figure}
%%%%%%%%%%%%%%%%%%%%%%%%%%%%%%%%%%%%%%%%%%%%%%%%%%%%%%%

In Fig.~\ref{JLST:gog} are displayed the partial waves ${\mathcal V}(^{2S+1}L_J)$ as calculated from several Gogny interactions and compared to the BHF results. The triplet spin combinations $^3S_1$-$^3D_1$ and $^3P_2$-$^3F_2$ have been displayed, since it is the BHF input at our disposal. The disagreements among $(S,T)$ channels, shown in Fig.~\ref{fig:STchannel} is translated, even magnified, for the partial waves. These results illustrates again that HF with an effective interaction can satisfactorily reproduce the BHF EoS as a sum of terms which are separately in disagreement. 
At the HF level, the $(S,T)$ channels are solely determined by the central and density-dependent terms of the interaction. The spin-triplet partial waves include also tensor and spin-orbit contributions. 
By inspecting their explicit expressions given in \ref{app:aux} one can see that for fixed values of $L$ and $S$, the contribution of central terms to ${\mathcal V}(^{2S+1}L_J)$ is the same but for a factor $(2J+1)$. In other words, some specific combinations as
%%%
\begin{eqnarray}\label{deltaP}
\delta_P&=&\frac{1}{3}\mathcal{V}_G(^{3}P_1)-\mathcal{V}_G(^{3}P_0) \;,\\
\label{deltaD}
\delta_D&=&\frac{1}{5}\mathcal{V}(^{3}D_2)-\frac{1}{7}\mathcal{V}_G(^{3}D_3)\;,  \\
\label{deltaF}
\delta_F&=&\frac{1}{7}\mathcal{V}_G(^{3}F_3)-\frac{1}{9}\mathcal{V}_G(^{3}F_4)\;,
\end{eqnarray}
allow us to isolate the tensor and the spin-orbit contributions. In principle, some other combinations are possible, but such combinations are contaminated by extra parity mixing which is absent in the present HF formalism. We will therefore restrict ourselves to the above ones.

It is worth stressing that this simple extraction of tensor and spin-orbit contributions as given in Eqs.~(\ref{deltaP}-\ref{deltaF}) is only possible for our {\it effective} interaction calculations. Indeed, the expectation value of the potential energy is done with plane-waves in the Fermi sea, and thus the contribution of each interaction term can be identified analytically, as shown in \ref{app:aux}. By inspecting these equations, we notice that apart from the different spatial form factors between Gogny and M3Y, the major difference is the presence of a finite-range spin-orbit term which thus contributes to all spin-triplet partial waves starting from $L=1$.
In the BHF case, or more generally in other calculations going beyond simple HF~\cite{hol72,hol10}, the extraction of a pure tensor and spin-orbit contribution to the EoS is not so simple as explained previously for the ($S,T$) decomposition. Since we are working with effective interactions, it is not necessary to separate all these contributions explicitly: we can absorb them (at least partially) directly into our effective coupling constants.

%%%%%%%%%%%%%%%%%%%%%%%%%%%%%%%%%%%%%%%%%%%%%%%%%%%%%%
\begin{figure}[h] 
 \centering
      \includegraphics[angle=-90,width=0.80\textwidth]{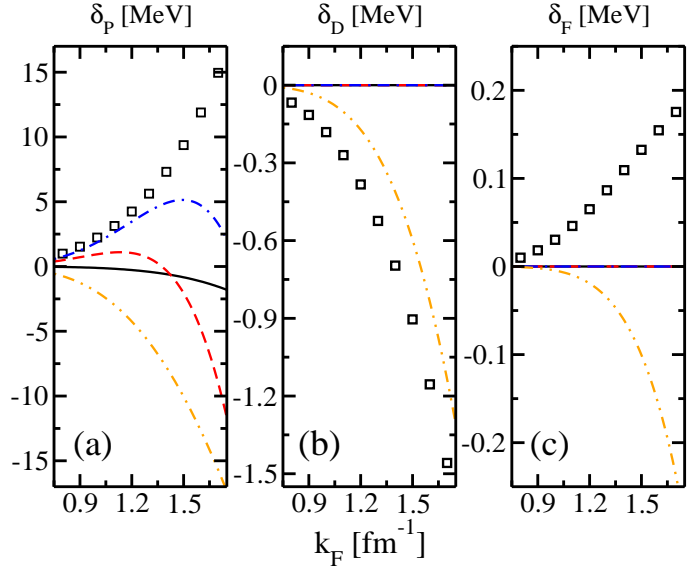}
   \caption{(Colors online) Difference of partial waves, expressed in MeV,  for BHF results (dots) and Gogny interactions (lines). The legend is the same as in Fig.\ref{JLST:gog}.}
   \label{deltas:gog}
\end{figure}
%%%%%%%%%%%%%%%%%%%%%%%%%%%%%%%%%%%%%%%%%%%%%%%%%%%%%%%

In Fig.~\ref{deltas:gog}, we compare the different Gogny parametrizations, Eqs.~(\ref{deltaP}-\ref{deltaF}) . For D1S, D1N and D1M, the splitting of the different partial waves with same $L$ arises only from spin-orbit term (see Eqs.~\ref{spin1:gog}). Since it has a zero-range nature, it acts only in the $L=1$ channel and the splitting of higher order waves as $D,F$ is zero. A different behaviour is observed when an explicit tensor term is added as in the D1ST2a case. In this case the splitting between partial waves is present also for the $D$ and $F$ waves although the sign is not correct for the $P$ and $F$ waves:  as already outlined in Ref.~\cite{Dav16PRC}, there is a possible inconsistency between the tensor parameters fitted using finite-nuclei constraints \cite{Gra13} and those deduced from infinite nuclear matter properties.This inconsistency was also emphasized in a previous analysis Refs~\cite{Pas14L,prep} based on Landau parameters, where a discrepancy between the sign of $H_l$ parameters and {\it ab-initio} results was spotted.

%%%%%%%%%%%%%%%%%%%%%%%%%%%%%%%%%%%%%%%%%%%%%%%%%%%%%%%
\begin{figure*}[h]
   \centering
      \includegraphics[angle=-90,width=0.45\textwidth]{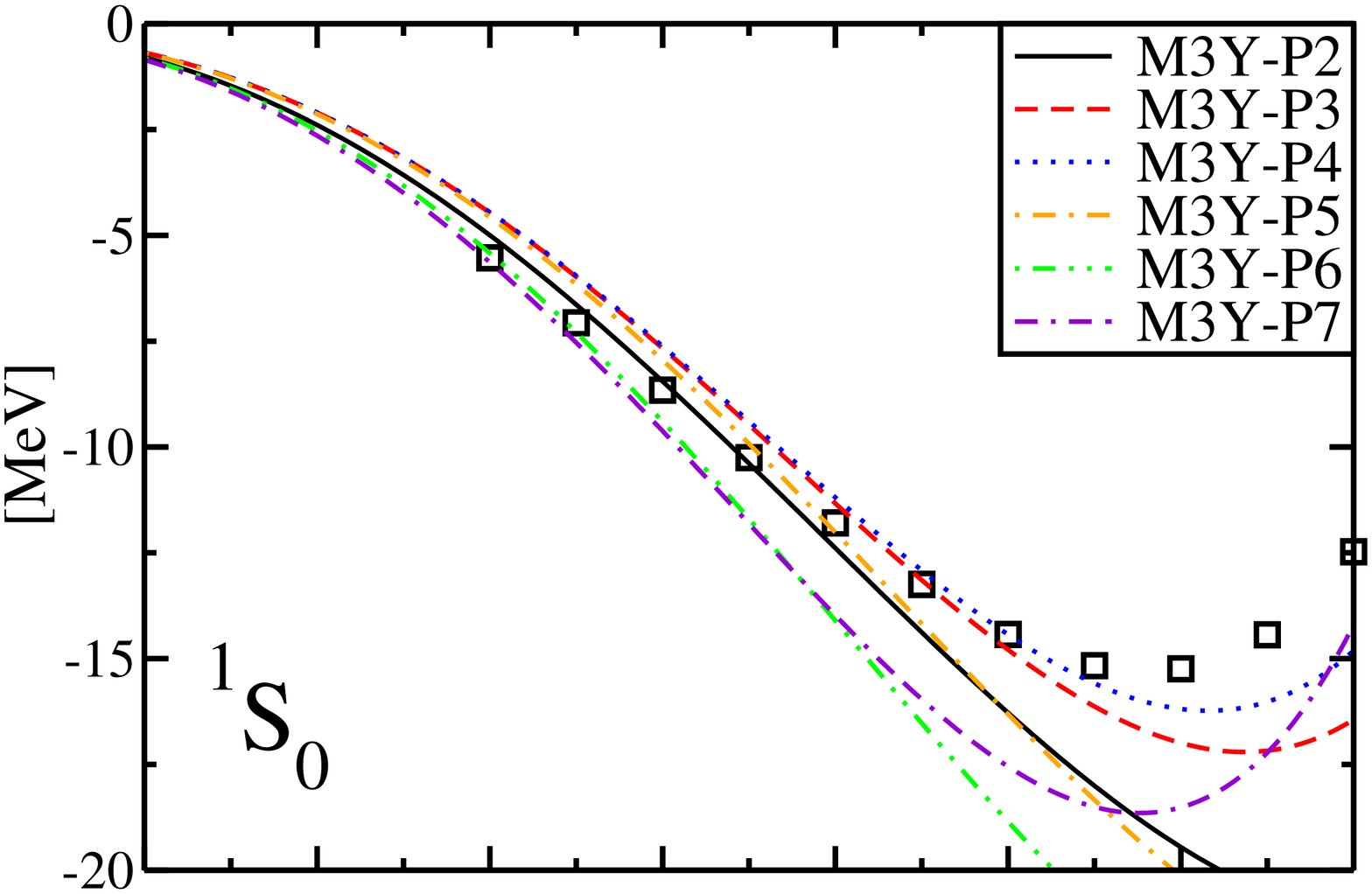}
      \hspace{-15.50mm}
            \includegraphics[angle=-90,width=0.45\textwidth]{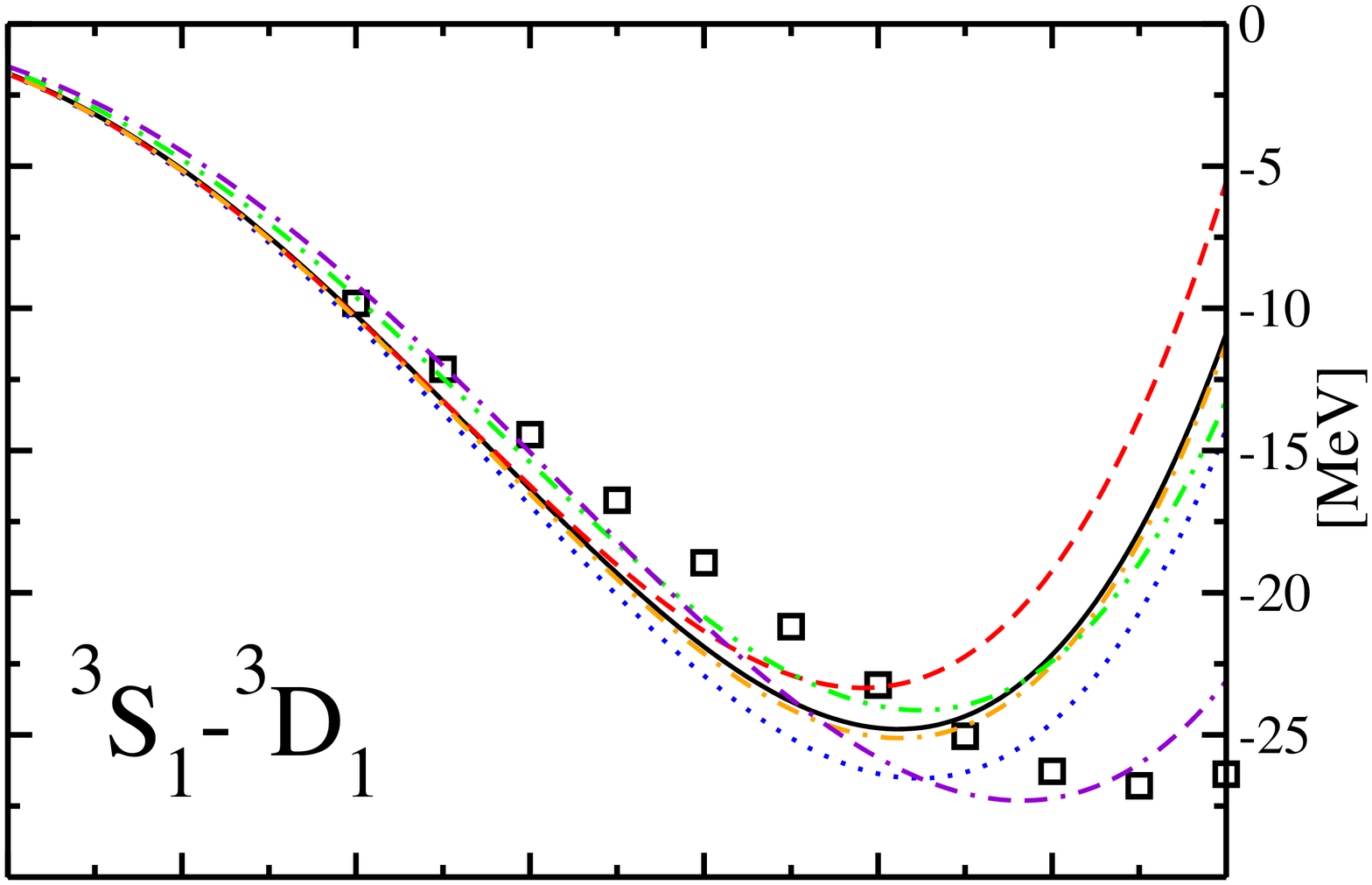}\\
             \vspace{-14.5mm}
                  \includegraphics[angle=-90,width=0.45\textwidth]{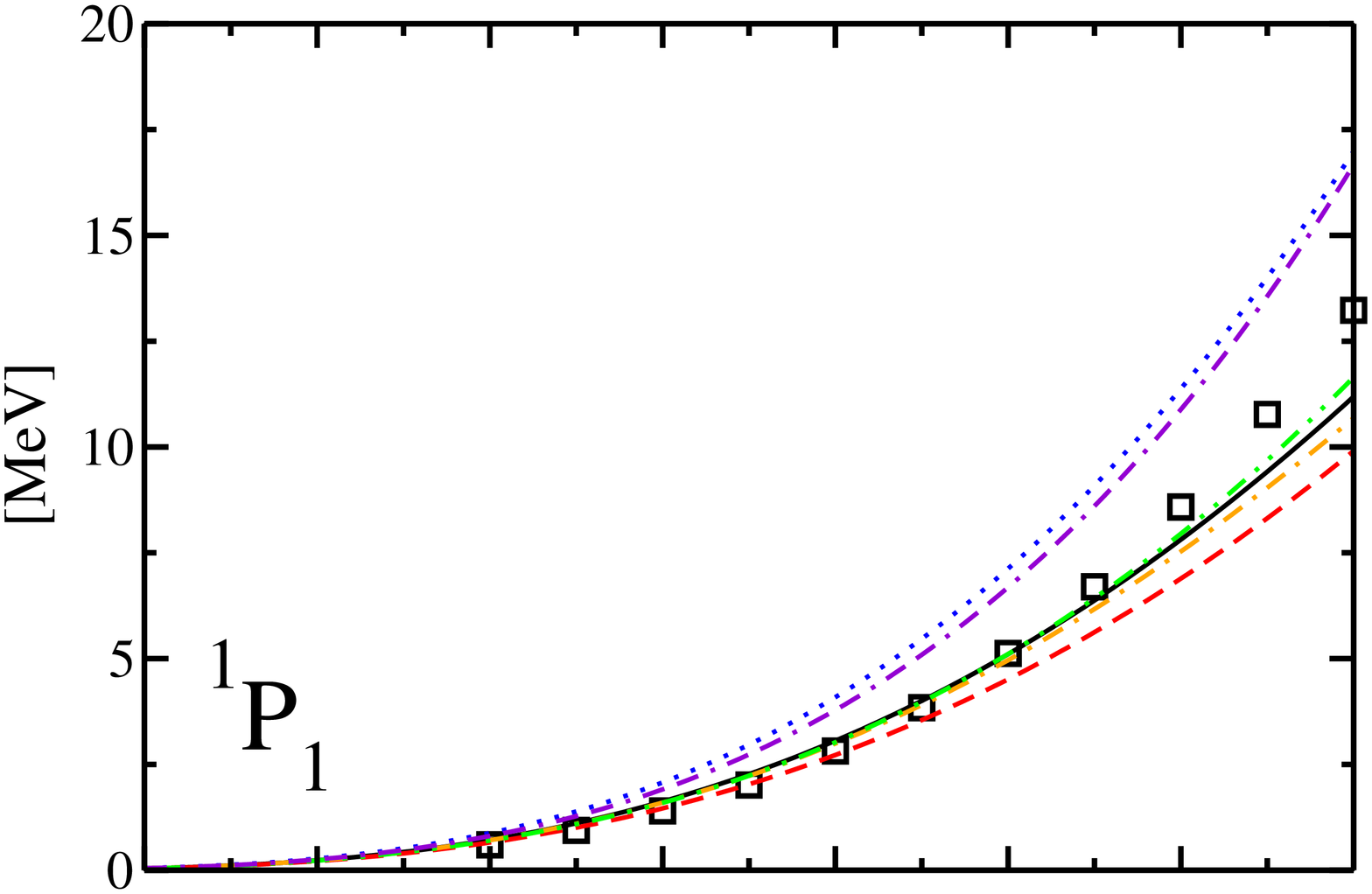}
      \hspace{-15.50mm}
                  \includegraphics[angle=-90,width=0.45\textwidth]{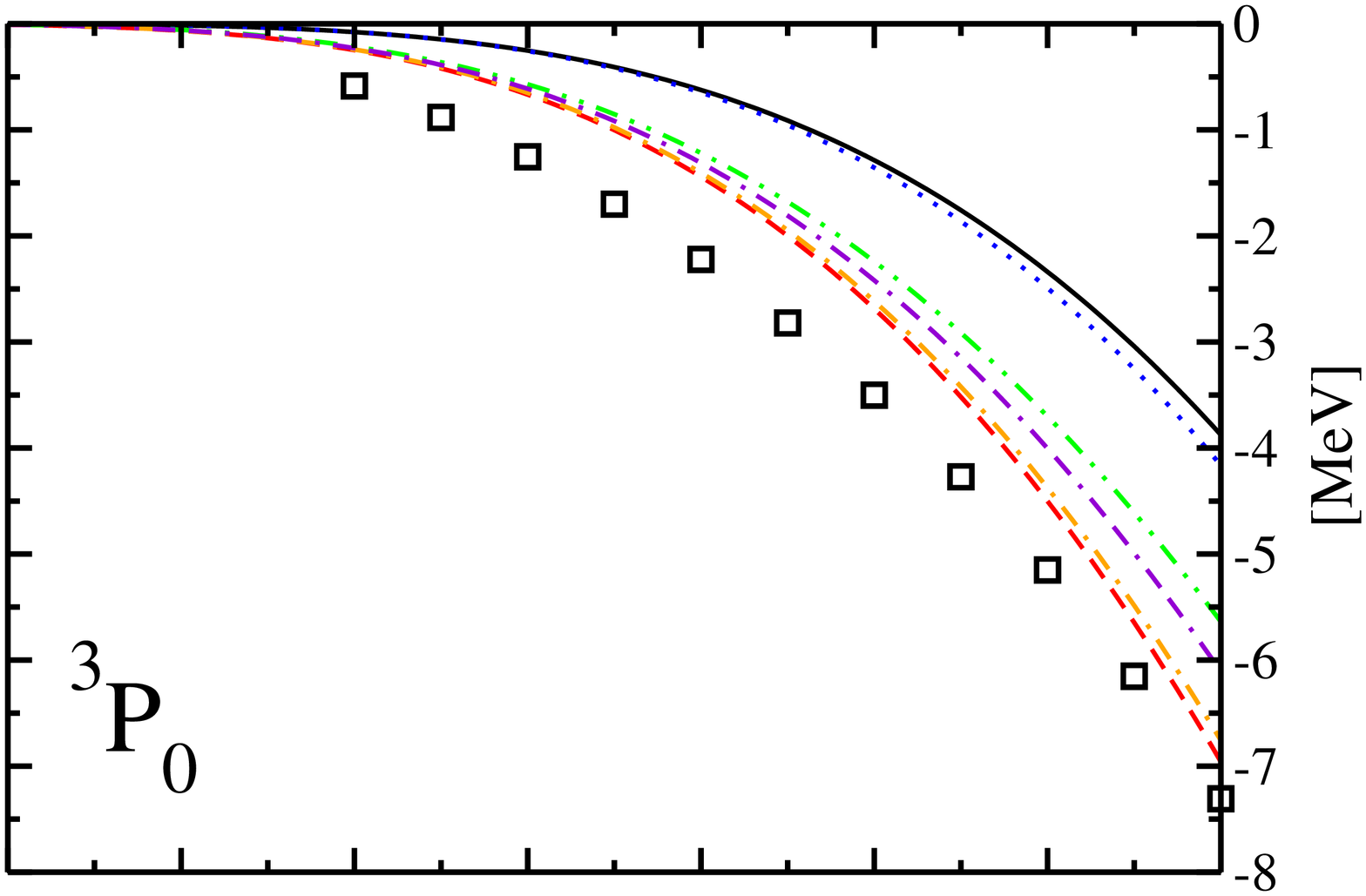}\\
             \vspace{-14.5mm}
                  \includegraphics[angle=-90,width=0.45\textwidth]{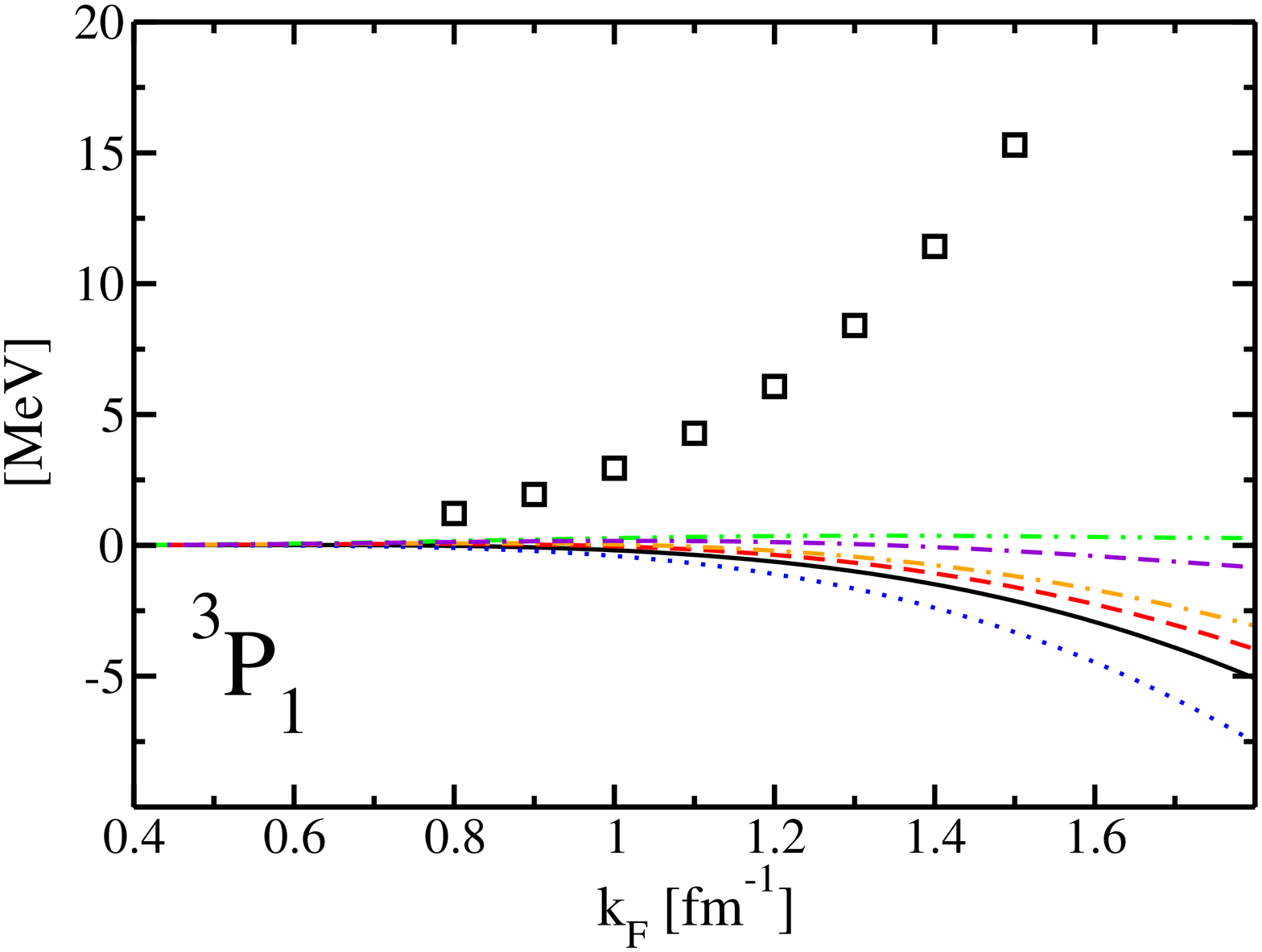}
      \hspace{-15.50mm}
                  \includegraphics[angle=-90,width=0.45\textwidth]{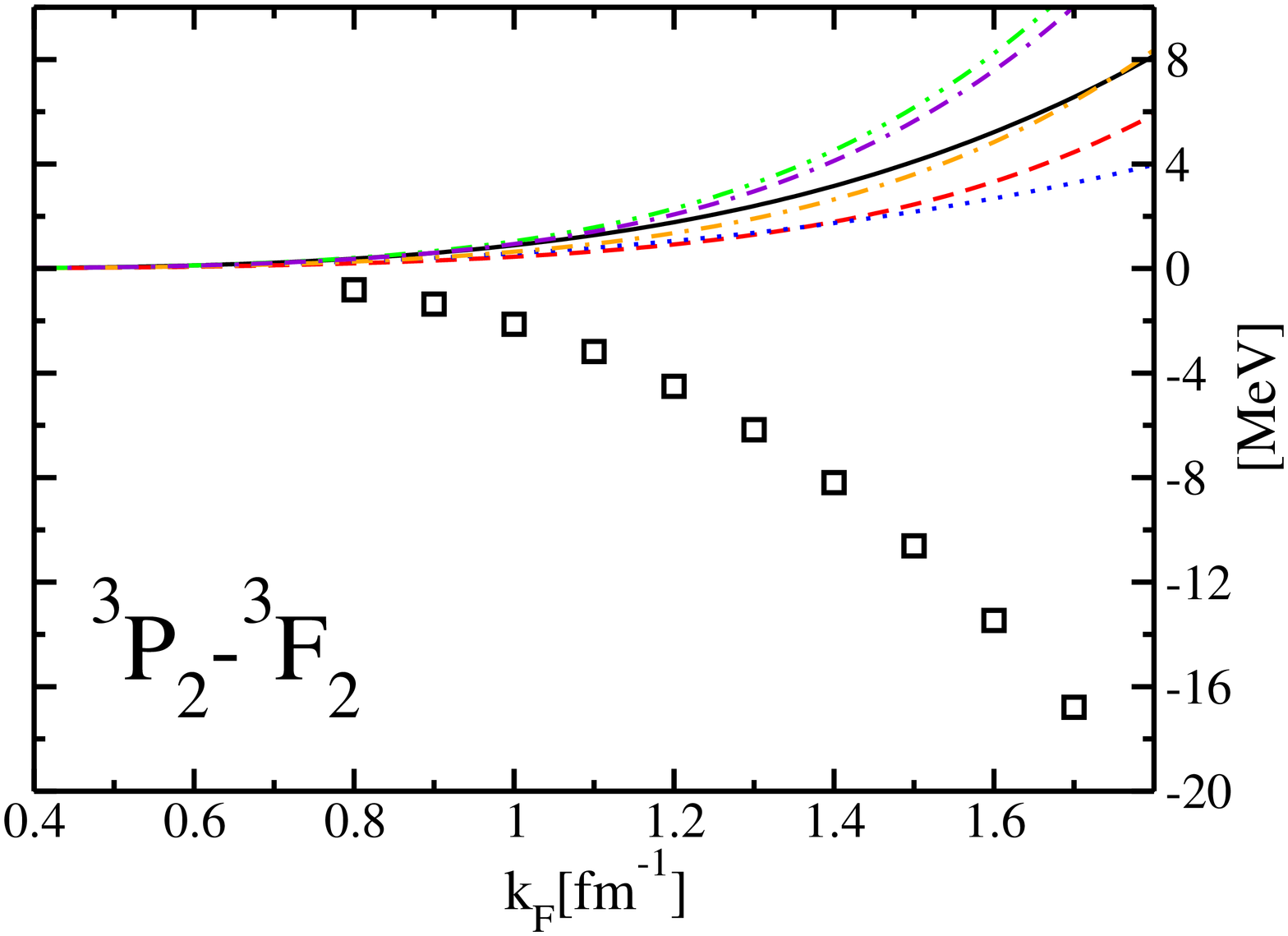}
   \caption{(Colors online) Partial wave decomposition of EoS in SNM for the different M3Y interactions (solid lines) and the BHF calculations (symbols)~\cite{bal97}.}
              \label{JLST}%
\end{figure*}
%%%%%%%%%%%%%%%%%%%%%%%%%%%%%%%%%%%%%%%%%%%%%%%%%%%%%%%

\begin{figure}[H]
 \centering
      \includegraphics[angle=-90,width=0.80\textwidth]{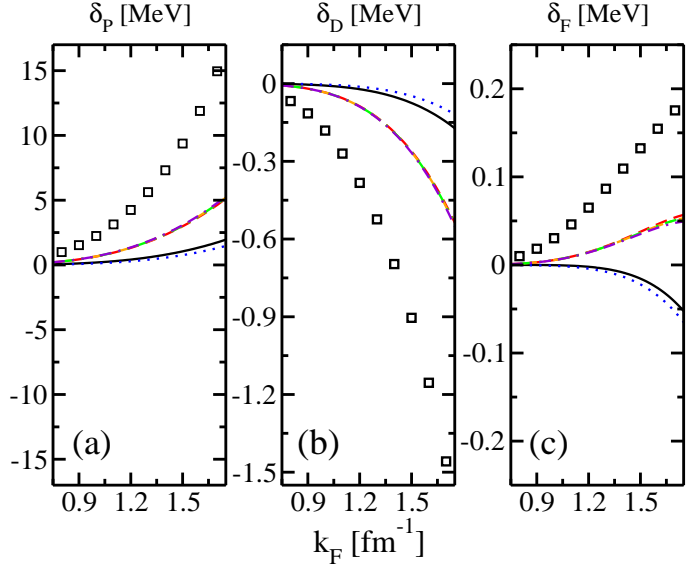}
   \caption{(Colors online) Difference of partial waves for BHF results (dots) and M3Y interactions (lines). The legend is the same as in Fig.\ref{JLST}.}
    \label{deltas:M3Y}
\end{figure}

The analogous analysis for the M3Y interaction is plotted in Figs.~\ref{JLST}-\ref{deltas:M3Y}, where we compare the partial wave decomposition of the M3Y-P$n$ with $n=2 \dots 7$ interactions with the BHF results given in Ref.~\cite{bal97} for some specific partial waves. We observe that the reproduction of the different channels $(J,L,S,T)$ is quite poor in the sense that some partial waves have the wrong sign. However the results are far better than the Gogny interactions and it is worth remarking that the $^1S_0$ channel, related to the pairing properties of the interaction, is globally well reproduced. Similarly to the previous part, we can isolate the spin-orbit and tensor contributions by using the differences between partial waves indicated in Eqs.~(\ref{deltaP}-\ref{deltaF}). We observe that the $P$ and $D$ channels have the correct sign, although the magnitude seems to be too low compared to BHF results. The results in the $F$ channels have the correct sign in the last generation of M3Y interactions. Such an observation is compatible with a previous analysis based on Landau parameters which appear to be too small compared to the results obtained with other {\it ab-initio} methods~\cite{Pas14L}.

%%%%%%%%%%%%%%%%%%%%%%%%%%%%%%%%%%%%%%%%%%%%%%%%%%%%%%
%%%%%%%%%%%%%%%%%%%%%%%%%%%%%%%%%%%%%%%%%%%%%%%%%%%%%%%

\section{From finite-range to zero-range}
\label{sect:finitezero}
In this section, we explore the zero-range limit of the M3Y and Gogny finite-range interactions. In particular, we follow the original Skyrme idea~\cite{sky59}, \emph{i.e.} performing a low-momentum expansion of the finite-range interaction. 
Actually, it has been shown~\cite{car08}, that through a density matrix expansion one can convert a finite-range effective interaction into a quasi-local functional \cite{car10,dob10}. Since only $L\le3$ partial waves matter for the EoS, and also for the nucleon-nucleon interaction itself, we can limit ourselves to order 6 in the momentum expansion, that is, to the N3LO pseudo-potential. For completeness we give in \ref{app:aux} the contributions for the different partial waves obtained with N3LO. These expressions correct some misprints in those given in \cite{Dav15}. 
To start with, we examine the general constraint imposed by the continuity equation and the local gauge invariance on N3LO.

\subsection{Local gauge invariance}
In the present HF context, the wave function is a Slater determinant built up from single-particle wave functions $ \psi_i (\mathbf{r})$, where index $i$ refers to the quantum numbers including the spin and isospin ones. These are eigenfunctions of the single-particle Hamiltonian $h$, which includes the mean field. 
In a general local gauge transformation, the wave functions are transformed as 
\begin{equation}
\psi_i (\mathbf{r}) \to e^{-i\phi(\mathbf{r})} \, \psi_i (\mathbf{r}) \;,
\end{equation}
where $\phi(\mathbf{r})$ is an arbitrary real function. The density matrix $\langle \mathbf{r}| \rho | \mathbf{r'} \rangle
= \sum_i \psi^\dagger _i(\mathbf{r}) \psi_i (\mathbf{r'})$ is transformed as
\begin{eqnarray*}
\langle \mathbf{r}| \rho | \mathbf{r'} \rangle \to
{\rm e}^{i(\phi(\mathbf{r})-\phi(\mathbf{r'}))}  \langle \mathbf{r}| \rho | \mathbf{r'} \rangle \simeq 
 \langle \mathbf{r}| \rho | \mathbf{r'} \rangle (1-i\phi(\mathbf{r'})+i\phi(\mathbf{r}))\;,
\end{eqnarray*}
where the last equation is valid at first order for an infinitesimal transformation.
Gauge invariance is imposed in a different way on the energy functional density
and on the pseudo-potential \cite{car08,rai11}. Consider for instance a standard Skyrme density functional, depending on the  the particle $\rho$, kinetic energy $\tau$ and current $\mathbf{j}$ densities, defined as 
\begin{eqnarray}
\rho (\mathbf{r})  &=& \sum_i \psi^\dagger _i(\mathbf{r}) \psi_i (\mathbf{r})\;, \\
\tau (\mathbf{r})  &=& \sum_i \vec \nabla \psi^\dagger _i(\mathbf{r}) \cdot \vec \nabla \psi_i (\mathbf{r})\;, \\
\mathbf{j} (\mathbf{r}) &=& \frac{\hbar}{2 m i} \sum_i \left[ \psi^\dagger_i (\mathbf{r}) \vec \nabla  \psi_i (\mathbf{r}) -  \vec \nabla \psi^\dagger_i (\mathbf{r})  \psi_i (\mathbf{r}) \right] \;.
\end{eqnarray}
It is easy to deduce that  under a gauge transformation the kinetic energy and current densities transform as $\tau \to \tau + 2 \mathbf{j} \nabla \phi + \rho (\nabla \phi)^2$ and $\mathbf{j} \to \mathbf{j} +\rho \nabla \phi$.  Any EDF of this type should contain the special combination $\rho \tau - \mathbf{j}^2$ to be gauge invariant. Actually, for the standard Skyrme potential, the gauge invariance is equivalent to the Galilean invariance and is thus satisfied~\cite{car08}. However, they are no longer equivalent when higher momentum powers are contained in the pseudo-potential~\cite{rai11}. Gauge invariance is much more constraining, and it has been imposed in order to reduce the number of terms of the general pseudo-potential at a given order in momenta~\cite{car08,rai11,dav13}. In fact, the problem of gauge invariance is linked to the problem of the continuity equation~\cite{bla86}, as we shall show in the following.

The continuity equation establishes the conservation of probability as a relation between the time variation of the particle density and the spatial variation of the current density
\begin{equation}\label{eq:cont}
\frac{\partial}{\partial t}\langle \rho \rangle + \text{div} \langle \mathbf{j} \rangle = 0\;.
\end{equation}
Let us apply an infinitesimal gauge transformation to the energy functional
\begin{eqnarray}\label{eq:Etilde}
E[\rho] & \to & E[\rho] + \int  d^3\mathbf{r} \; d^3\mathbf{r'} \frac{\delta E}{\delta \langle \mathbf{r}| \rho | \mathbf{r'} \rangle} \delta \langle \mathbf{r}| \rho | \mathbf{r'} \rangle \nnn
                    & \simeq & E[\rho] + i \int  d^3\mathbf{r} \; d^3\mathbf{r'} \langle \mathbf{r'}| h | \mathbf{r} \rangle \langle \mathbf{r}| \rho | \mathbf{r'} \rangle (-\phi(\mathbf{r'})+\phi(\mathbf{r})) \nnn
                    & = & E[\rho] - i \int  d^3\mathbf{r} \; \langle \mathbf{r}| [h,\rho] | \mathbf{r} \rangle \phi(\mathbf{r})\nnn                   
                    & = & E[\rho] + \hbar \int  d^3\mathbf{r} \; \frac{\partial}{\partial t} \langle \rho \rangle \phi(\mathbf{r})\;.
\end{eqnarray}
In the same way, the kinetic energy term transforms as
\begin{eqnarray}\label{eq:Ttilde}
T & \to & \frac{\hbar^2}{2 m}  \sum_i \int  d^3\mathbf{r} \; \vec \nabla\psi^\dagger_i (\mathbf{r}) \vec \nabla  \psi_i (\mathbf{r}) \nnn
           & = & T - \hbar \int  d^3\mathbf{r} \; \phi (\mathbf{r}) \vec \nabla \langle \mathbf{j} (\mathbf{r}) \rangle \;.
\end{eqnarray}
A direct comparison between Eqs.~(\ref{eq:Etilde}) and (\ref{eq:Ttilde}) clearly shows that in order to preserve the continuity equation Eq.~(\ref{eq:cont}), the potential has to be gauge invariant. This invariance is trivially fulfilled for a local potential, and becomes relevant for momentum dependent pseudo-potentials, as is the Skyrme one and its N$\ell$LO generalizations. The continuity equation is satisfied by the standard Skyrme (N1LO) interaction, as was explicitly checked by Engel {\it et al.}~\cite{eng75}. For higher orders, the situation is different and the local gauge invariance induces some strong constraints on the possible combinations entering in the potential. Actually the only compatible form for N3LO is the one given in Eqs. (\ref{eq:N3LO:c}) and (\ref{eq:N3LO:t}) for the central and tensor parts respectively~\cite{rai11b,dav13}. In order to see how this potential can be useful, we now examine the zero-range limit of our finite-range potentials.

\subsection{M3Y}

We begin by considering the central term of the M3Y interaction, which is the simplest one. In momentum space it reads
%%%
\begin{eqnarray}\label{central:kk}
 v^C_{N}(\mathbf{k},\mathbf{k}') = \sum_{n=1}^3 \left[ t_n^{(SE)}P_{SE} +t_n^{(TE)}P_{TE}+t_n^{(SO)}P_{SO}+t_n^{(TO)}P_{TO}  \right] \, 
  \frac{1}{\mu^{C}_n(\mathbf{k}-\mathbf{k}')^2+(\mu^{C}_n)^3} .
 \end{eqnarray}
 %%%
The momentum expansion gives
%%%
\begin{eqnarray}\label{eq:vcen|:exp}
v^C_{N}(\mathbf{k},\mathbf{k}')&\approx& \sum_{n=1}^3 \frac{C_n}{(\mu^{C}_n)^3}-  \frac{C_n}{(\mu^{C}_n)^5}    \big[ \bk^{\prime 2}  + \bk^2 +2 \bkp \cdot  \bk \big]    \, \nonumber\\
 & + &  \frac{C_n}{(\mu^{C}_n)^7} \left[({\mathbf{k}}^2 + {\mathbf{k}'}^2)^2 + 4 ({\mathbf{k}'} \cdot {\mathbf{k}})^2-4({\bf k'} \cdot {\mathbf{k}}) ({\mathbf{k}}^2 + {\mathbf{k}'}^2) \right]\nonumber\\
 & - & \frac{C_n}{(\mu^{C}_n)^9} \left\{ ({\mathbf{k}'}^{2} + {\mathbf{k}}^{2})\left[ ({\mathbf{k}'}^{2}+{\mathbf{k}}^{2})^{2}+12({\mathbf{k}'}\cdot{\mathbf{k}})^{2}\right] 
- 2({\mathbf{k}'} \cdot {\mathbf{k}}) \left[3({\mathbf{k}'}^{2}+{\mathbf{k}}^{2})^{2}+4({\mathbf{k}'}\cdot{\mathbf{k}})^{2}\right]\right\} \nnn
&+& \dots ,
\end{eqnarray}
%%%
where we have defined the following combinations of parameters 
%%%
\begin{eqnarray}
4C_n&=& (t^{(SE)}_n+t^{(SO)}_n+t^{(TE)}_n+t^{(TO)}_n) +(t^{(SE)}_n-t^{(SO)}_n-t^{(TE)}_n+t^{(TO)}_n) P_{\tau} \nonumber\\
&& + (-t^{(SE)}_n-t^{(SO)}_n+t^{(TE)}_n+t^{(TO)}_n) P_{\sigma} 
 +(-t^{(SE)}_n+t^{(SO)}_n-t^{(TE)}_n+t^{(TO)}_n)P_{\sigma}P_{\tau} 
\end{eqnarray}
%%%
We immediately recognise the form of the central part of  N3LO pseudo-potential as given in Eq.~(\ref{eq:N3LO:c}). Even if, due to the explicit presence of the isospin exchange operator $P_\tau$, we can not map directly on the operator form M3Y parameters with those of N3LO, we see that there are some specific relations between the coefficients $t_1^{(n)}$ and $t_2^{(n)}$. More precisely, by comparing the momentum dependence of Eqs.~(\ref{eq:vcen|:exp}) and (\ref{eq:N3LO:c}), we observe that  Eq.~(\ref{eq:vcen|:exp}) can be considered as a special case of the N3LO pseudo-potential when $t_1^{(n)} = -t_2^{(n)}$, \dots We can thus argue that N3LO contains more freedom in the sense that, at this order of expansion, Eq.~(\ref{eq:vcen|:exp}) is more restrictive. 

Actually, this is quite general: any finite-range interaction leads to a N3LO-like pseudo-potential with relations between the coefficients. This can be seen by a formal Taylor expansion of a scalar function $F(\mathbf{k} - \mathbf{k}')$. Explicitly it reads
%%%
\begin{eqnarray}\label{eq.taylor}
F(\mathbf{k} - \mathbf{k}') & = & A_0 + A_2 (\mathbf{k} - \mathbf{k}')^2 + A_4 (\mathbf{k} - \mathbf{k}')^4 + \cdots  \nnn
& = & A_0 + A_2 \left[{\mathbf{k}}^2 + {\mathbf{k}'}^2 - 2{\mathbf{k}'} \cdot {\mathbf{k}}\right] \nnn
&& + A_4  \left[({\mathbf{k}}^2 + {\mathbf{k}'}^2)^2 + 4 ({\mathbf{k}'} \cdot {\mathbf{k}})^2 - 4({\mathbf{k}'} \cdot {\mathbf{k}})({\mathbf{k}}^2 + {\mathbf{k}'}^2) 
\right] + \cdots
\end{eqnarray}
%%%
which obviously reproduces the required structure.
We can apply the same strategy to the tensor term to obtain
%%%
\begin{eqnarray}\label{eq:vtenkk}
v^{T}_{N}(\mathbf{k},\mathbf{k}') = -\sum_n \left[t_n^{TNE}P_{TE}+t_n^{TNO}P_{TO} \right] \frac{32\pi}{\mu^{T}_n}
\frac{1}{\left[(\mu^{T}_n)^2+(\mathbf{k}-\mathbf{k}')^2\right]^3}\left[T_e(\mathbf{k},\mathbf{k}')-T_o(\mathbf{k},\mathbf{k}') \right] ,
\end{eqnarray}
%%%
which leads to
%%%
\begin{eqnarray}
v^{T}_{N}(\mathbf{k},\mathbf{k}')&\approx&-\sum_n \left[(t_n^{TNE}+t_n^{TNO})+(-t_n^{TNE}+t_n^{TNO})P_{\tau} 
+(t_n^{TNE}+t_n^{TNO})P_{\sigma}+(-t_n^{TNE}+t_n^{TNO})P_{\sigma}P_{\tau} \right] \nonumber\\
&\times&\frac{8\pi}{(\mu^{T}_n)^7} \bigg\{ 
\left( T_e(\mathbf{k},\mathbf{k}')-T_o(\mathbf{k},\mathbf{k}') \right)  \nonumber\\
&& \quad -\frac{3}{(\mu^{T}_n)^2}\left[ (\mathbf{k}^2+\mathbf{k}^{'2}) T_e(\mathbf{k},\mathbf{k}')+2\mathbf{k}\cdot \mathbf{k}' T_o(\mathbf{k},\mathbf{k}')\right]  \nnn
&& \quad +\frac{3}{(\mu^{T}_n)^2}\left[ (\mathbf{k}^2+\mathbf{k}^{'2}) T_o(\mathbf{k},\mathbf{k}')+2\mathbf{k}\cdot \mathbf{k}' T_e(\mathbf{k},\mathbf{k}')\right]  \nonumber\\
&& \quad +\frac{6}{(\mu^{T}_n)^4}\left[ \left(\frac{1}{4}(\mathbf{k}^2+\mathbf{k}^{'2})^2+\frac{1}{8} (\mathbf{k}\cdot \mathbf{k}')^2 \right)T_e(\mathbf{k},\mathbf{k}') 
 +2(\mathbf{k}\cdot \mathbf{k}' )(\mathbf{k}^2+\mathbf{k}^{'2})T_o(\mathbf{k},\mathbf{k}')\right]  \nonumber\\
&& \quad -\frac{6}{(\mu^{T}_n)^4} \left[ \left(\frac{1}{4}(\mathbf{k}^2+\mathbf{k}^{'2})^2+\frac{1}{8} (\mathbf{k}\cdot \mathbf{k}')^2 \right)T_o(\mathbf{k},\mathbf{k}')  
+2(\mathbf{k}\cdot \mathbf{k}' )(\mathbf{k}^2+\mathbf{k}^{'2})T_e(\mathbf{k},\mathbf{k}') \right] \nonumber \\
&& \quad +\dots  \bigg\} \;.
\end{eqnarray}
%%%
Analogously to the central part, we observe that this expansion has the same structure as the N3LO pseudo-potential. A direct comparison between the above equation and Eq.~(\ref{eq:N3LO:t}) shows that the N3LO parameters deduced from the expansion are again not independent. 

Let us finally consider the spin-orbit term. Its expression in momentum space takes the form
\begin{eqnarray}\label{eq:soN}
 v^{SO}_{N}(\mathbf{k},\mathbf{k}') = 8\pi i  \sum_{n=1}^2\left[ t^{LSE}_nP_{TE} + t^{LSO}_nP_{TO} \right](\mathbf{k}\times\mathbf{k}')\cdot (\mathbf{s}_1+\mathbf{s}_2) 
  \frac{1}{\mu^{SO}_n \left[(\mathbf{k}-\mathbf{k}')^2+(\mu^{so}_n)^2 \right]^2}\;.
\end{eqnarray}
%
%\noindent For the spin-orbit term a similar Taylor expansion can be done. 
%By considering  only the first order of its Taylor expansion we find standard momentum dependence of spin-orbit term in Skyrme interactions~\cite{bel59}
%\begin{eqnarray*}\label{central:kk:so}
%v^{so}_{N}(\mathbf{k},\mathbf{k}') &\approx& \frac{2\pi i}{(\mu^{so}_n)^5}  \sum_{n=1}^2\left[(t_n^{LNE}+t_n^{LNO})+(t_n^{LNO}-t_n^{LNE})P_{\tau} +(t_n^{LNE}+t_n^{LNO})P_{\sigma} \right. \nnn
%&& \left. \hspace{2cm} +(t_n^{LNO}-t_n^{LNE})P_{\sigma}P_{\tau} \right] (\mathbf{k}\times\mathbf{k}')\cdot (\mathbf{s}_1+\mathbf{s}_2)\;.
% \end{eqnarray*}
%The higher-order terms in this expansion implies different combinations of momentum and spin operator~\cite{rai11}. To be more specific we obtain (only the $k$-dependence is written for the sake of clarity)
%\bea\label{so:exp}
%V(\vec k,\vec k') & = & \frac{i}{\mu^{so}_n \left[(\mathbf{k}-\mathbf{k}')^2+(\mu^{so}_n)^2 \right]^2}(\mathbf{k'} \times \mathbf{k})\cdot(\vec \sigma_1+\vec \sigma_2)\nnn 
%& \simeq &  \frac{i}{(\mu^{so}_n)^5} (\mathbf{k'} \times \mathbf{k}) \cdot(\vec \sigma_1+\vec \sigma_2) \; \left[1-\frac{2}{(\mu^{so}_n)^2}(\mathbf{k'}^2 +\mathbf{k}^2-2 \mathbf{k'} \cdot \mathbf{k})+ ...\right] \nn 
%\eea
%where the last line corresponds to the momentum expansion. 

Up to second order Taylor expansion it reads
\begin{eqnarray}\label{so:exp}
v^{SO}_{N}(\mathbf{k},\mathbf{k}') &\approx& \frac{2\pi i}{(\mu^{SO}_n)^5}  \sum_{n=1}^2\left[(t_n^{LNE}+t_n^{LNO})+(t_n^{LNO}-t_n^{LNE})P_{\tau} +(t_n^{LNE}+t_n^{LNO})P_{\sigma} \right. \nnn
&& \left. \hspace{2cm} +(t_n^{LNO}-t_n^{LNE})P_{\sigma}P_{\tau} \right] (\mathbf{k}\times\mathbf{k}')\cdot (\mathbf{s}_1+\mathbf{s}_2) \nnn
&&  \hspace{3cm} \left[1-\frac{2}{(\mu^{SO}_n)^2}(\mathbf{k'}^2 +\mathbf{k}^2-2 \mathbf{k'} \cdot \mathbf{k})+ ...\right] 
 \end{eqnarray}
Such an expansion has already been considered in Refs.~\cite{car08,rai11}. The authors have shown that terms like $(\mathbf{k'}^2 +\mathbf{k}^2) \; \left[ (\mathbf{k'} \times \mathbf{k})\cdot(\vec \sigma_1+\vec \sigma_2)\right]$ or $ \mathbf{k'} \cdot \mathbf{k} \; \left[(\mathbf{k'} \times \mathbf{k})\cdot(\vec \sigma_1+\vec \sigma_2)\right]$ are not gauge invariant (separately or in any combination). The same is true for higher order terms. Actually, only the term $(\mathbf{k'} \times \mathbf{k})\cdot(\vec \sigma_1+\vec \sigma_2)$, \emph{i.e.} the zero-range spin-orbit term, is gauge invariant, and it  
has the same form as the standard spin-orbit term in Skyrme interactions~\cite{bel59}. 
In Ref.~\cite{rai11b}, the authors have proved a one-to-one relation between local gauge invariance and continuity equation as given in Eq.~(\ref{eq:cont}).

Since the momentum expansion in Eq.~(\ref{so:exp})  contains terms which violate the local gauge invariance. We conclude that the finite-range spin-orbit term violates the continuity equation, which  is only compatible with a zero-range spin-orbit interaction. However, even if rigorously a finite-range spin-orbit term seems forbidden, the question of its validity and use from a phenomenological point of view remains open since to our knowledge, there is no study to estimate to which extent the continuity equation is violated in that case. 

\subsection{Gogny interaction}

\begin{figure}[h]
 \centering
      \includegraphics[angle=-90,width=0.48\textwidth]{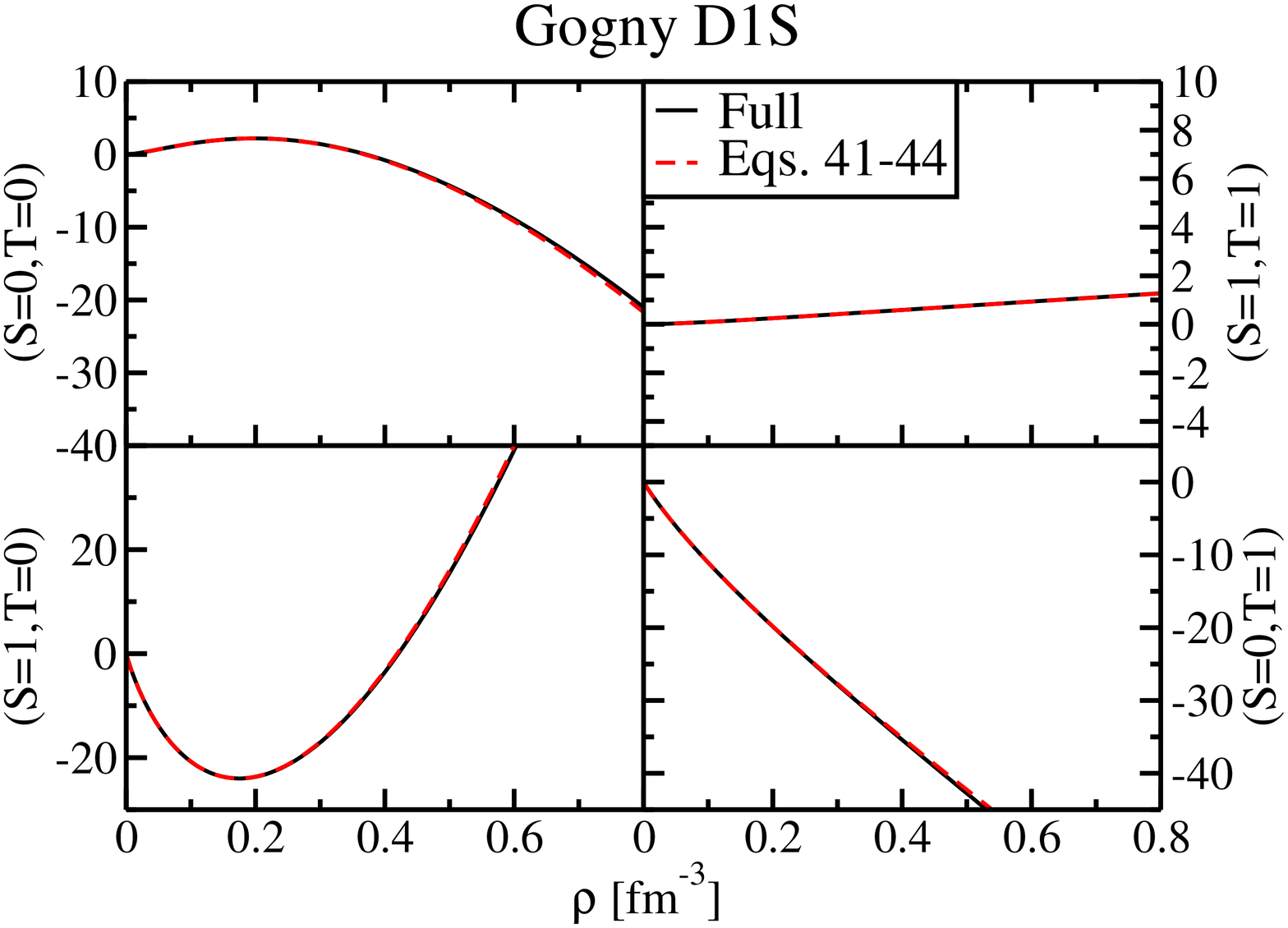}
            \includegraphics[angle=-90,width=0.48\textwidth]{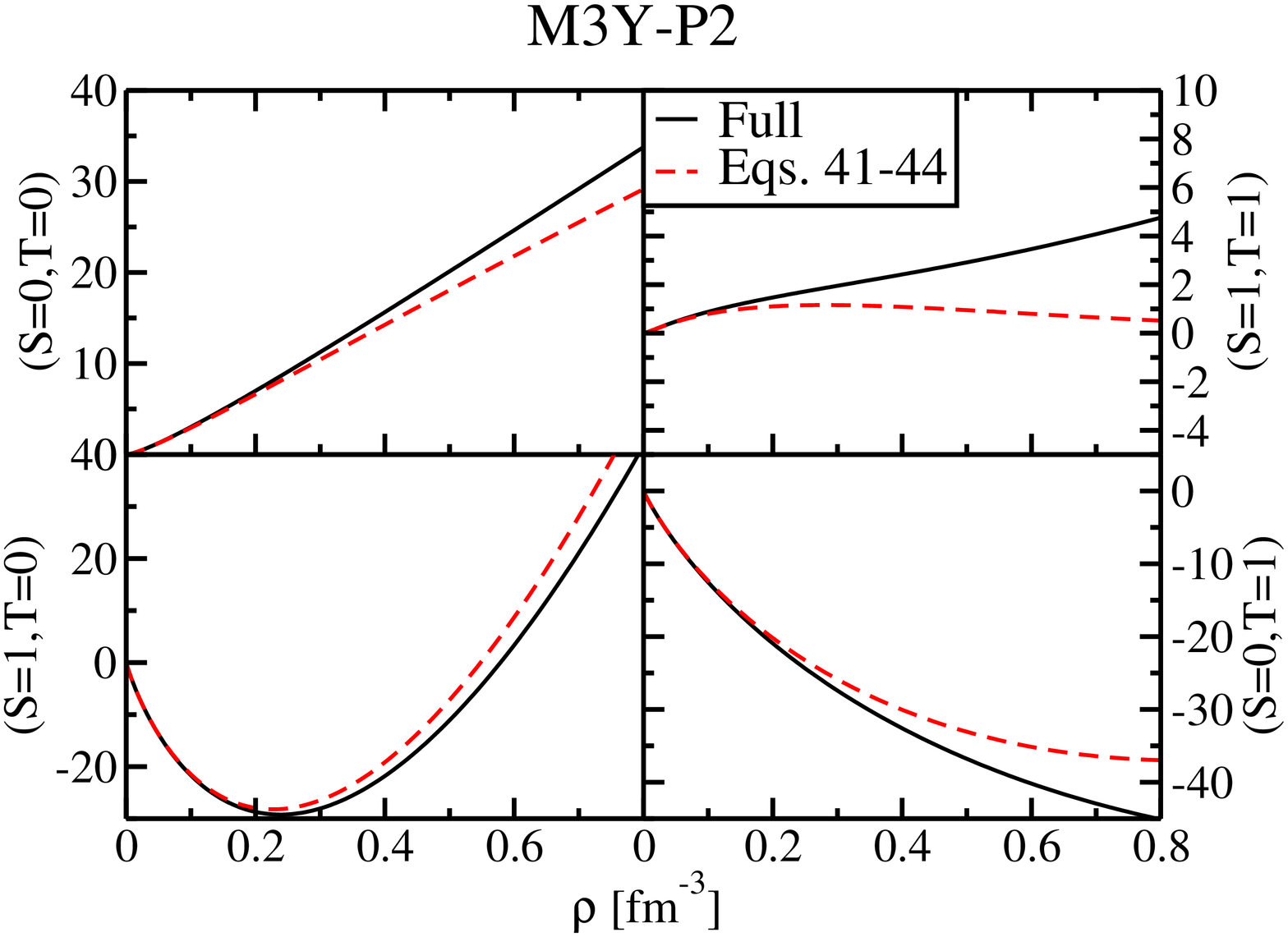}
   \caption{(Colors online). $(S ,T )$ channels of the EoS, expressed in MeV, obtained with the Gogny D1S interaction (left panel) and the M3Y-P2 interaction (right panel).}
    \label{ST:recomposed}
\end{figure}

We study now the momentum expansion of a Gogny  interaction. In momentum space, the central and the tensor terms are written as
%%%
\begin{eqnarray}\label{eq:cenG}
 v^C_{G}(\mathbf{k},\mathbf{k}') = \sum_{n=1}^2 \left[ W^{n}+B^{(n)}P_{\sigma}-H^{(n)}P_{\tau}-M^{(n)}P_{\sigma}P_{\tau}  \right] 
  \pi ^{3/2} (\mu^{C}_n)^3 e^{-(\mathbf{k}-\mathbf{k}')^2/4(\mu^{C}_n)^2} , 
 \end{eqnarray}
 %%%
and 
%%%
\begin{eqnarray}
v^{T}_{G}(\mathbf{k},\mathbf{k}') =  -\frac{\pi^{3/2}(\mu^{T})^7}{4} (V_{T1}+V_{T2}P_{\tau})e^{-(\mathbf{k}-\mathbf{k}')^2/4(\mu^{T})^2} 
 \left[T_e(\mathbf{k},\mathbf{k}')-T_o(\mathbf{k},\mathbf{k}') \right] .
\end{eqnarray}
%%%
Expanding the form factor  $e^{-(\mathbf{k}-\mathbf{k}')^2/4\mu^2} $ using Eq.~(\ref{eq.taylor}), we obtain the same conclusion as for the M3Y interactions: we find a N3LO-like structure and the deduced N3LO interaction parameters are related to each other.  We can then conclude that the N3LO pseudo-potential is actually the common limit of all finite-range central and tensor interactions. The essential point is now to understand to which extent this limit is reliable, that is, contains the main physical ingredients of the original interaction. A first indication has been given in Fig.~\ref{Eos:convergence2}, where we have shown that beyond $L=3$, the contributions to the global EoS were very small. We can go one step further by considering the following equations which give, for N3LO, the content of each $(S,T)$ channels in terms of partial waves
%%%
\bea
\mathcal{V}^{(0,0)} &=&\mathcal{V}(^{1}P_1)+\mathcal{V}(^{1}F_3)   \label{eq:reconstructionA},\\
\mathcal{V}^{(0,1)} &=&\mathcal{V}(^{1}S_0)+\mathcal{V}(^{1}D_2) , \label{eq:reconstructionB}\\
\mathcal{V}^{(1,0)} &=&\mathcal{V}(^{3}S_1)+\mathcal{V}(^{3}D_1)+\mathcal{V}(^{3}D_2)+\mathcal{V}(^{3}D_3)   , \label{eq:reconstructionC}\\
\mathcal{V}^{(1,1)} &=&\mathcal{V}(^{3}P_0)+\mathcal{V}(^{3}P_1)+\mathcal{V}(^{3}P_2)+\mathcal{V}(^{3}F_2)+\mathcal{V}(^{3}F_3)+\mathcal{V}(^{3}F_4) . \label{eq:reconstructionD}
\eea
%%%
These equations do not hold for a finite-range interaction since in this case, a given channel should receive an infinite number of different contributions. However we have checked that these contributions coming from partial waves with $L>3$ do not contribute significantly to the $(S,T)$ channels.
In Fig.\ref{ST:recomposed} we compare the results of the $(S,T)$-channels obtained with the complete interaction, $i.e.$ considering \emph{all} partial waves and restricting ourselves up to $L=3$ wave as shown in Eqs.~(\ref{eq:reconstructionA}-\ref{eq:reconstructionD}).

%%%%%%%%%%%%%%%%%%%%%%%%%%%%%%%%%%%%%%%%%%%%%%%%%%%%%%
%%%%%%%%%%%%%%%%%%%%%%%%%%%%%%%%%%%%%%%%%%%%%%%%%%%%%%%

From this figure, we observe that the properties of the EoS are very well reproduced up to saturation density, by simply considering partial  waves with $L\le3$. This means that the 6th-order expansion described above, $i.e.$ the induced N3LO pseudo-potential, constitutes a very good approximation of \emph{any} initial finite-range potential. We can thus argue that the general N3LO pseudo-potential, that is without any specific relation between its coefficients, certainly contains enough degrees of freedom to be an excellent quantitative substitute of any reasonable finite-range interaction used in nuclear structure calculations.
We notice that the Gaussian form factor in Gogny D1S essentially gets no contributions from partial waves larger than L=3 on a very large density range. On the contrary, the Yukawa form factor has a much slower convergence rate and we observe that above $\rho\approx0.2$ fm$^{-3}$ higher order partial waves could still play a role, this is particularly evident in the $(S=T=1)$ channel. This is not really surprising and it is related to the convergence properties of the Taylor expansion of the Yukawa potential.

\section{Conclusions}\label{sect:conclusions}
 
We have presented in detail the different contributions to the infinite matter EoS of two widely used finite-range interactions, namely Gogny and M3Y. The method presented here is actually general and can be applied to any other finite-range interaction with different form factors. In particular, we have shown how to extract more informations concerning spin-orbit and tensor terms of the interaction, by using a coupled basis, either to spin and isospin $(S,T)$ channels or 
to partial waves $(J,L,S,T)$, which includes also an explicit coupling to angular and total angular momentum.  
We stress that BHF results have been used here to illustrate our method, but other microscopic methods can also be employed, provided  they include $(S,T)$ and $(J,L,S,T)$ results in a wide density range. 

The existing Gogny and M3Y parameterizations give a fair description of the EoS in SNM, as given by BHF calculations, even at values of density beyond the saturation point. However, by looking at the EoS decomposition in $(S,T)$ channels 
we have shown that this satisfactory result is obtained by adding up four terms which can notoriously differ from BHF ones, with important differences between these interactions. Whereas the M3Y family is globally in reasonable agreement with BHF $(S,T)$ channels, the Gogny one does not reproduce them beyond the saturation density, with even a qualitative disagreement. We have shown that the addition of a third Gaussian to the central component gives enough freedom to properly reproduce all $(S,T)$ channels even at large values of the density. The spin-orbit and tensor components of the interaction can be tested by looking at $(J,L,S,T)$ partial waves. The present M3Y parameterizations are in qualitative agreement with BHF calculations only for the $^1S_0$, $^3S_1$-$^3D_1$, $^1P_1$ and $^3P_0$ components up to the saturation density. We suggest this analysis is of particular interest to determine more precise parameterizations, as the fit of such quantities could provide a reasonable starting point for a complete fit in finite nuclei~\cite{kor13}.

A momentum expansion of both Gogny and M3Y interactions up to the sixth order shows that the resulting series matches exactly with the N3LO Skyrme pseudo-potential derived in Ref.~\cite{rai11,dav13}, thus proving 
the original idea of Skyrme~\cite{sky59}, not only for the central term, but also for the tensor one.
By doing that, we have also found that the finite-range spin-orbit can be expanded on a series of terms which are not gauge-invariant, apart from the first one, which represents the standard Skyrme spin-orbit term~\cite{cha97}. 
This point was already signaled~\cite{dav13} for the extensions suggested by Skyrme. Therefore, the finite range spin-orbit as used in the M3Y interaction is in conflict with the continuity equation~\cite{rai11}. We have not quantified the violation neither guessed its impact on finite nuclei calculations and we thus expect further investigations along this direction.
To study the convergence of such an expansion, we have used the partial wave analysis to study which are the dominant partial waves entering the EoS and its decomposition in ($S,T$) channels. We have found that the Gogny interaction gets contributions essentially from partial waves up to $L=3$ in a very large range of densities. The role of the higher order partial waves being negligible. This is also true for the Yukawa potential, but only around saturation density.
The conclusion of such a study is that the N3LO pseudo-potential contains enough physics to replace \emph{any} finite-range potential, within a Hartree-Fock context, and to properly describe the physics around saturation density, which is the region of interest for calculations in finite nuclei.
Such encouraging results are in agreement with previous studies based on Density Matrix Expansion~\cite{car10}. 

\section*{Acknowledgments} 

We thank M. Baldo for kindly providing us with his BHF results. We also thank A. Rios for valuable suggestions on the manuscript. The work of J.N. has been supported by grant  FIS2014-51948-C2-1-P, Mineco (Spain). 

%%%%%%%%%%%%%%%%%%%%%%%%%%%%%%%%%%%%%%%%%%%%%%
%% APPENDIX
%%%%%%%%%%%%%%%%%%%%%%%%%%%%%%%%%%%%%%%%%%%%%%

\appendix

\section{Symmetric nuclear matter and pure neutron matter potential energies}~\label{app:EOS}
For completeness, we recall here the results for the symmetric nuclear matter and pure neutron matter EoS for the different interactions considered in the text: Gogny, M3Y and Skyrme N3LO. In the following, $k_F (k_{F_n})$ represents the SNM (neutron) Fermi momentum, and $\rho\;(\rho_n)$ the SNM (neutron) density. 
\subsection*{Gogny interaction}
%%%
\bea
 {\cal V}^{SNM} &=&  \sum_{i=1}^2 \frac{4W_i+2B_i -2 H_i - M_i}{12\sqrt{\pi}} (k_F \mu^{C}_i)^3 \nnn
&& 
- \sum_{i=1}^2 \frac{W_i+2B_i -2 H_i - 4 M_i}{2\sqrt{\pi}} G(k_F \mu^{C}_i)  
+ \frac{3}{8} t^{(DD)} \rho^{\alpha+1} \nnn
 {\cal V}^{PNM} &=& \sum_{i=1}^2\frac{2W_i+B_i -2 H_i -M_i}{12\sqrt{\pi}} (k_{Fn} \mu^{C}_i)^3 \;,\nnn
&& -\sum_{i=1}^2\frac{W_i+2B_i-H_i-2M_i}{2\sqrt{\pi}} G(k_{Fn}\mu^{C}_i) 
+\frac{1}{4}t^{(DD)} (1-x^{(DD)})\rho_n^{\alpha+1} \;,\nn
\end{eqnarray}
%%%
 with
\bea
G(a) =  \frac{1}{a^3}  \bigg[ {\rm e}^{-a^2}\left[a^2-2\right]  + 2-3a^2   +\sqrt{\pi}a^3 \hbox{Erf} (a)\bigg] , \nn
\eea
where $\text{Erf}(x)$ is the error function (see Ref~\cite{abr64} by instance).
\subsection*{M3Y interaction}
%%%
\begin{eqnarray}
 {\cal V}^{SNM} &=& \sum_{i=1}^3 \frac{3t^{(SE)}_i+t^{(SO)}_i+3t^{(TE)}_i+9t^{(TO)}_i}{12\pi} \left(\frac{k_F}{\mu^{C}_i}\right)^3 \nnn
& & 
 -\sum_{i=1}^3\frac{-3t^{(SE)}_i+t^{(SO)}_i-3t^{(TE)}_i+9t^{(TO)}_i}{4\pi}Y\left(\frac{k_F}{\mu^{C}_i}\right)
 + \frac{3}{8}t^{(dd)}\rho^{\alpha+1}\;, \nnn
 {\cal V}^{PNM} & = &\sum_{i=1}^3\frac{t^{(SE)}_i+3t^{(TO)}_i}{6\pi}\left(\frac{k_{F}}{\mu^{C}_i}\right)^3 %\nnn
 %& &  
 -\sum_{i=1}^3\frac{-t^{(SE)}_i+3t^{(TO)}_i}{2\pi}Y\left(\frac{k_F}{\mu^{c}_i}\right) 
 +\frac{1}{4}t^{(DD)} \left[1-x^{(DD)}\right]\rho^{\alpha+1}\;, \nn
\end{eqnarray}
%%%
with
\bea
Y(a) = \frac{1}{32 a^3} \bigg[ 4 a^2 \left( 6 a^2-1\right) -32 a^3\arctan(2 a)+(1+12 a^2)\hbox{ln}(1 + 4 a^2) \bigg] . \nn
\eea
\subsection*{N3LO interaction}

\begin{eqnarray}
 {\cal V}^{SNM} &=&\frac{3}{8}t_0\rho+\frac{3}{80} a^2\left[ 3t_1+(5+4x_2)t_2\right] \rho^{5/3}+\frac{9}{280}a^4\left[ 3t^{(4)}_1+(5+4x^{(4)}_2)t^{(4)}_2\right] \rho^{7/3}\nnn
 &+&\frac{2}{15}a^6\left[ 3t^{(6)}_1+(5+4x^{(6)}_2)t^{(6)}_2\right] \rho^3 + \frac{1}{16}t_3\rho^{\alpha+1}\;, \nnn
 {\cal V}^{PNM} & = &\frac{1}{4}(1-x_0)t_0\rho_n+\frac{3}{40} b^2\left[ t_1 (1-x_1)+3 t_2(1+x_2)\right] \rho^{5/3}_n \nnn
 &+& \frac{9}{140} b^4\left[ t^{(4)}_1 (1-x^{(4)}_1)+3 t^{(4)}_2(1+x^{(4)}_2)\right] \rho^{7/3}_n\nnn
 &+& \frac{4}{15} b^6 \left[ t^{(6)}_1 (1-x^{(6)}_1)+3 t^{(6)}_2(1+x^{(6)}_2)\right] \rho^{3}_n+\frac{1}{24}t_3 (1-x_3)\rho_n^{\alpha+1} \;,\nn
\end{eqnarray}

\noindent where $a=\left(3\pi^2/2 \right)^{1/3}$ and $b=\left(3\pi^2 \right)^{1/3}$.

\section{Contributions to the $(S,T)$ channels in symmetric nuclear matter}~\label{app:ST}
We present here the analytical results for the four spin-isospin channels for the different  interactions.
\subsection*{Gogny interaction}
%%%
\bea
\mathcal{V}^{(0,0)}  &=&    \sum_{i=1}^2 \, (W_i - B_i  + H_i  - M_i) \, G^{(0,0)}(k_F \mu^{C}_i)\;,  \nnn
\mathcal{V}^{(0,1)}  &=& 3 \sum_{i=1}^2 \, (W_i - B_i  - H_i + M_i) \, G^{(0,1)}(k_F \mu^{C}_i)
   + \frac{3}{16} t^{(DD)}(1-x^{(DD)}) \rho^{\alpha+1}  \;,\nnn
\mathcal{V}^{(1,0)} &=&   3 \sum_{i=1}^2 \, (W_i + B_i + H_i + M_i) \, G^{(1,0)}(k_F \mu^{C}_i)
    + \frac{3}{16} t^{(DD)}(1+x^{(DD)}) \rho^{\alpha+1}\;, \nnn
\mathcal{V}^{(1,1)} &=& 9  \sum_{i=1}^2 \, (W_i + B_i - H_i - M_i) \, G^{(1,1)}(k_F \mu^{C}_i)\;, \nn
\eea
%%%
with
\bea
G^{(S,T)}(a)  =    \frac{1}{48 \sqrt{\pi}}  \bigg[  a^3  - 6 \, (-1)^{S+T} G(a) \bigg]\,. \nn
\eea
The function $G(a)$ has been defined in~\ref{app:EOS}.  

\subsection*{M3Y interaction}
%%%
\bea
\mathcal{V}^{(0,0)}  &=&  \sum_{n} \, t^{(SO)}_n  \, Y^{(0,0)}\left(\frac{k_F}{\mu^{C}_n}\right) , \nnn
\mathcal{V}^{(0,1)}  &=& 3  \, \sum_{n} \,  t^{(SE)}_n  \, Y^{(0,1)}\left(\frac{k_F}{\mu^{C}_n}\right) 
+ \frac{3}{16} t^{(DD)}(1-x^{(DD)}) \rho^{\alpha+1} , \nnn
\mathcal{V}^{(1,0)}  &=& 3  \, \sum_{n} \,  t^{(TE)}_n \, Y^{(1,0)}\left(\frac{k_F}{\mu^{C}_n}\right) 
+ \frac{3}{16} t^{(DD)}(1+x^{(DD)}) \rho^{\alpha+1} , \nnn
\mathcal{V}^{(1,1)}  &=& 9  \,  \sum_{n} \, t^{(TO)}_n  \, Y^{(1,1)}\left(\frac{k_F}{\mu^{C}_n}\right) , \nn
   \eea
   %%%
where
\bea
Y^{(S,T)}(a) = \frac{1}{12 \pi}   \bigg[ a^3 - 3\, (-1)^{S+T} Y(a) \bigg]\,. \nn
\eea
The function $Y(a)$ has been defined in~\ref{app:EOS}.

\subsection*{N3LO}
The N3LO pseudo-potential is obtained as a momentum expansion up to sixth-order in gradients~\cite{car08,rai11}. 
The expressions for the $(S,T)$ channels in SNM read
%%%
\bea
\mathcal{V}^{(0,0)} &=&
 \frac{3}{160} t_2 (1-x_2)  \rho k_F^2 
+ \frac{9}{560} t_2^{(4)} (1-x_2^{(4)})  \rho k_F^4  
 + \frac{1}{15} t_2^{(6)} (1-x_2^{(6)})  \rho k_F^6 \,,\nnn
%%%
\mathcal{V}^{(0,1)} &=&   \frac{3}{16} t_0(1-x_0) \rho
+ \frac{1}{32} t_3(1-x_3) \rho^{\alpha+1}
+\frac{9}{160} t_1 (1-x_1) \rho k_F^2   \nnn
&&  +  \frac{27}{560} t_1^{(4)} (1- x_1^{(4)}) \rho k_F^4  
+ \frac{1}{5} t_1^{(6)} (1-x_1^{(6)} ) \rho  k_F^6  
\,,\nnn
%%%
\mathcal{V}^{(1,0)} &=&  \frac{3}{16} t_0(1+x_0)  \rho
+ \frac{1}{32}  t_3(1+x_3)  \rho k_F^{3(\alpha+1)}  +\frac{9}{160} t_1 (1+x_1)  \rho k_F^2   \nnn
&&  +  \frac{27}{560} t_1^{(4)} (1+ x_1^{(4)})  \rho k_F^4 
+ \frac{1}{5} t_1^{(6)} (1+x_1^{(6)} )   \rho k_F^6
\,,\nnn
%%%
\mathcal{V}^{(1,1)} &=& 
 \frac{27}{160} t_2 (1+x_2)   \rho k_F^2
+ \frac{81}{560} t_2^{(4)} (1+x_2^{(4)} )  \rho k_F^4 +\frac{3}{5} t_2^{(6)} (1+x_2^{(6)})  \rho k_F^6 \, . \nn
\eea
%%%

Adding up all these $(S,T)$ contributions for each type of interaction, we recover the respective results for the global EoS presented in~\ref{app:EOS}.

\mbox{} \vfill \eject

\section{Partial wave decomposition in symmetric nuclear matter}~\label{app:aux}
\subsection*{Gogny interaction}

%In the following, we include in our expressions explicitly the contribution of a tensor as described in Ref.~\cite{Ang11}
\begin{eqnarray}
\mathcal{V}(^1S_0) & = & 3\sum_{n=1}^2 \left[W^{(n)}-B^{(n)}-H^{(n)}+M^{(n)} \right]G_0^{(C)}(k_F \mu^{C}_n) \;,\nnn
\mathcal{V}(^1P_1) & = & 3\sum_{n=1}^2 \left[W^{(n)}-B^{(n)}+H^{(n)}-M^{(n)} \right]G_1^{(C)}(k_F \mu^{C}_n)\;, \nnn
\mathcal{V}(^1D_2) & = & 15\sum_{n=1}^2 \left[W^{(n)}-B^{(n)}-H^{(n)}+M^{(n)} \right]G_2^{(C)}(k_F \mu^{C}_n) \;,\nnn
\mathcal{V}(^1F_3) & = & 7\sum_{n=1}^2 \left[W^{(n)}-B^{(n)}+H^{(n)}-M^{(n)} \right]G_3^{(C)}(k_F \mu^{C}_n) \;,\nn
\end{eqnarray}
%%%
\begin{eqnarray}\label{spin1:gog}
\mathcal{V}(^3S_1) & = & 3\sum_{n=1}^2 \left[W^{(n)}+B^{(n)}+H^{(n)}+M^{(n)} \right]G_0^{(C)}(k_F \mu^{C}_n)+\frac{3}{8}t^{(DD)}\rho^{\alpha+1} \;,\nnn
\mathcal{V}(^3P_{J=0,1,2}) & = & 3(2J+1)\sum_{n=1}^2 \left[W^{(n)}+B^{(n)}-H^{(n)}-M^{(n)} \right]G_1^{(C)}(k_F \mu^{C}_n) \nnn
&& +\frac{W_0}{80}\rho k_F^2\left\{2\delta_{J,0}+3\delta_{J,1}-5\delta_{J,2} \right\} \nnn
&&  + (V_{T1}+V_{T2})(\mu^{T})^2G_1^{(T)}(k_F\mu^{T})\left\{2\delta_{J,0}-3\delta_{J,1}+\delta_{J,2} \right\} \;,\nnn
\mathcal{V}(^3D_{J=1,2,3}) & = & (2J+1)\sum_{n=1}^2 \left[W^{(n)}+B^{(n)}+H^{(n)}+M^{(n)} \right]G_2^{(C)}(k_F \mu^{C}_n)\nnn
& &  +(V_{T1}-V_{T2})(\mu^{T})^2G_2^{(T)}(k_F\mu^{T})\left\{3\delta_{J,1}-5\delta_{J,2}+2\delta_{J,3} \right\}\;, \nnn
\mathcal{V}(^3F_{J=2,3,4}) & = & 3(2J+1)\sum_{n=1}^2 \left[W^{(n)}+B^{(n)}-H^{(n)}-M^{(n)} \right]G_3^{(C)}(k_F \mu^{C}_n)\nnn
& &  +(V_{T1}+V_{T2})(\mu^{T})^2G_3^{(T)}(k_F \mu^{T})\left\{3\delta_{J,1}-5\delta_{J,2}+2\delta_{J,3} \right\}\;. \nn
\end{eqnarray}
The symbol $\delta_{J,x}$ with $x=1,2,3\dots$ is the standard Kronecker delta.
The functions $G^{(C,T)}_{j=0,1,\dots}$ are defined as
%%%
\begin{eqnarray}
G^{(C)}_0(a)&=&\frac{1}{8\sqrt{\pi}a^3}\left\{ -2+6a^2+3a^4+2(1-2a^2)e^{-a^2}-4a^3\sqrt{\pi}{\text{Erf}}(a)\right\}\;, \nnn
G^{(C)}_1(a)&=&\frac{3}{8\sqrt{\pi}a^3}\left\{ 2+2a^2+a^4-2(1+2a^2)e^{-a^2}-4a^3\sqrt{\pi}\text{Erf}(a)+4a^2S(a)\right\}\;, \nnn
G^{(C)}_2(a)&=&\frac{1}{8\sqrt{\pi}a^3}\left\{ 26-30a^2+3a^4-(26+20a^2)e^{-a^2}-20a^3\sqrt{\pi}\text{Erf}(a) \right.  \nnn
&& \left. \hspace{1cm}+(24+36a^2)S(a)\right\} \;,\nnn
G^{(C)}_3(a)&=&\frac{1}{8\sqrt{\pi}a^5}\left\{ 72-14a^2-114a^4+3a^6-(72+58a^2+28a^4)e^{-a^2} \right. \nnn
&& \left. \hspace{1cm} -28a^5\sqrt{\pi}\text{Erf}(a)+a^2(120+72a^2)S(a)\right\}\;,\nn 
\end{eqnarray}
\begin{eqnarray}
G^{(T)}_1(a)&=&\frac{9}{40\sqrt{\pi}a^3}\left\{ 6-6a^2-a^4-6e^{-a^2}+4a^2S(a)\right\} \;,\nnn
G^{(T)}_2(a)&=&\frac{3}{40\sqrt{\pi}a^3}\left\{ 10-34a^2-a^4-10e^{-a^2}+12(2+a^2)S(a)\right\}\;, \nnn
G^{(T)}_3(a)&=&\frac{9}{40\sqrt{\pi}a^5}\left\{ 120-94a^2-86a^4-a^6-(120+26a^2)e^{-a^2} 
+24a^2(5+a^2)S(a)\right\}\;, \nn
\end{eqnarray}
with
\begin{eqnarray}
S(a)=\gamma-\text{Ei}(-a^2)+\ln(a^2)\;, \nn
\end{eqnarray}
$\gamma$ being the Euler gamma constant and Ei the exponential integral~\cite{abr64}.

\subsection*{M3Y  interaction}
\begin{eqnarray}
 \mathcal{V}(^1S_0)&=&\sum_n 3t_n^{(SE)}Y_0^{(C)}(k_F/\mu^{C}_n)
 +\frac{3}{16}t^{(DD)}(1-x^{(DD)})\rho^{\alpha+1} \;,\nnn
  \mathcal{V}(^1P_1)&=&\sum_n 3t_n^{(SO)}Y_1^{(C)}(k_F/\mu^{C}_n)\;, \nnn
  \mathcal{V}(^1D_2)&=&\sum_n 15t_n^{(SE)}Y_2^{(C)}(k_F/\mu^{C}_n) \;,\nnn
  \mathcal{V}(^1F_3)&=&\sum_n 7t_n^{(SE)}Y_3^{(C)}(k_F/\mu^{C}_n)\;, \nn
\end{eqnarray}
%%%
%%%
\begin{eqnarray}
 \mathcal{V}(^3S_1)&=&\sum_n 3t_n^{(TE)}Y_0^{(C)}(k_F/\mu^{C}_n)
 +\frac{3}{16}t^{(DD)}(1+x^{(DD)})\rho^{\alpha+1} \;,\nnn
 \mathcal{V}(^3P_{J=0,1,2})&=&\sum_n 3(2J+1)t_n^{(TO)}Y_1^{(C)}(k_F/\mu^{C}_n) \nnn
 && +\sum_n t^{(LSO)}_n Y_1^{(C)}(k_F/\mu^{SO}_n)\left\{ 2\delta_{J,0}+3\delta_{J,1}-5\delta_{J,2}\right\}\nnn
   & & + \sum_n \frac{9}{80\pi} \frac{t^{TNO}_n}{(\mu^{T}_n)^2}Y_1^{(T)}(k_F/\mu^{T}_n)\left\{ 2\delta_{J,0}-3\delta_{J,1}+\delta_{J,2}\right\} \;,\nnn
 \mathcal{V}(^3D_{J=1,2,3})&=&\sum_n(2J+1) t_n^{(TE)}Y_2^{(C)}(k_F/\mu^{C}_n) \nnn
 && +\frac{1}{3}\sum_n t^{(LSE)}_n Y_2^{(C)}(k_F/\mu^{SO}_n)\left\{ 9\delta_{J,1}+5\delta_{J,2}-14\delta_{J,3}\right\}\nnn
   & &  + \sum_n \frac{3}{80\pi} \frac{t^{(TNE)}_n}{(\mu^{T}_n)^2}Y_2^{(T)}(k_F/\mu^{T}_n)\left\{ 3\delta_{J,1}-5\delta_{J,1}+2\delta_{J,3}\right\}\;, \nnn
 \mathcal{V}(^3F_{J=2,3,4})&=&\sum_n 3(2J+1) t_n^{(TO)}Y_3^{(C)}(k_F/\mu^{C}_n) \nnn
 && +\sum_n t^{(LSO)}_n Y_3^{(C)}(k_F/\mu^{SO}_n)\left\{ 20\delta_{J,2}+7\delta_{J,3}-27\delta_{J,4}\right\}\nnn
    & &   + \sum_n \frac{9}{160\pi} \frac{t^{(TNO)}_n}{(\mu^{T}_n)^2}Y_3^{(T)}(k_F/\mu^{T}_n)\left\{ 4\delta_{J,2}-7\delta_{J,3}+3\delta_{J,4}\right\} \;.\nn
  \end{eqnarray}

\noindent The symbol $\delta_{J,x}$ with $x=1,2,3\dots$ is the standard Kronecker delta.
The functions $Y^{(C,T)}_{j=0,1,\dots}$ are defined as
%%%
\begin{eqnarray}
Y_0^{(C)}(a)&=&\frac{1}{128\pi a^3} \bigg\{ 4a^2-168a^4+128a^3 \arctan(2a)+(-1-24a^2+48a^4)\ln(1+4a^2) \bigg\} \;,\nnn
Y_1^{(C)}(a)&=&\frac{3}{128\pi a^3} \bigg\{ -4a^2-88a^4+128a^3 \arctan(2a) \nnn
&& \hspace{1cm} +(1-24a^2+16a^4)\ln(1+4a^2)+16a^2 \text{Li}_2(-4a^2) \bigg\} \;,\nnn
Y_2^{(C)}(a)&=&\frac{1}{128\pi a^3} \bigg\{ -172a^2-312a^4+640a^3\arctan(2a) \nnn
&& \hspace{1cm} +(31-24a^2+48a^4)\ln(1+4a^2) +12(-1+12a^2 )\text{Li}_2(-4a^2) \bigg\} \;,\nnn
Y_3^{(C)}(a)&=&\frac{1}{128\pi a^5} \bigg\{ 12a^2-596a^4-344a^6+896a^5\arctan(2a) \nnn
&& \hspace{1cm} +(-3+83a^2+168a^4+48a^6)\ln(1+4a^2) 
+ 12a^2(-5+24a^2 )\text{Li}_2(-4a^2) \bigg\} \;,\nn
\end{eqnarray}
%%%
%%%
\begin{eqnarray}
Y_1^{(T)}(a)&=&\frac{1}{a^3}\left\{ 4a^2(3-2a^2)-3(1+4a^2)\ln(1+4a^2)-8a^2\text{Li}_2(-4a^2)\right\}\;, \nnn
Y_2^{(T)}(a)&=&\frac{1}{a^3}\left\{ 4a^2(29-2a^2)-17(1+4a^2)\ln(1+4a^2)+12(1-2a^2)\text{Li}_2(-4a^2)\right\} \;,\nnn
Y_3^{(T)}(a)&=&\frac{1}{a^5}\left\{ -4a^2(15-176a^2+4a^4)+(15-26a^2-344a^4)\ln(1+4a^2) 
+24a^2(5-4a^2)\text{Li}_2(-4a^2)\right\}. \nn
\end{eqnarray}
The $\text{Li}_2(z)$ is the polylogarithmic function defined in Eq.27.7.1 of Ref.~\cite{abr64}. See also Ref.~\cite{mor79}. 

\subsection*{N3LO interaction}
%%% 
The N3LO Skyrme pseudo-potential is obtained by taking into account all possible combinations of gradients and exchange operators up to 6th order~\cite{rai11}. By construction, the only partial waves included are the $S,P,D,F$ waves. The $JLST$ decomposition of the potential energy reads
\begin{eqnarray}
\mathcal{V}(^{1}S_0)&=&\frac{3}{16}t_{0}(1-x_{0})\rho+\frac{1}{32}t_{3}(1-x_{3})\rho^{\alpha+1}+\frac{9}{160}t_{1}(1-x_{1})\rho k_{F}^{2} \nonumber\\
& & +\frac{9}{280}t_{1}^{(4)}(1-x_{1}^{(4)})\rho k_{F}^{4}+\frac{1}{10}t_{1}^{(6)}(1-x_{1}^{(6)})\rho k_{F}^{6}\,,\nnn
\mathcal{V}(^{3}S_1)&=&\frac{3}{16}t_{0}(1+x_{0})\rho+\frac{1}{32}t_{3}(1+x_{3})\rho^{\alpha+1}+\frac{9}{160}t_{1}(1+x_{1})\rho k_{F}^{2}\nonumber\\
& & +\frac{9}{280}t_{1}^{(4)}(1+x_{1}^{(4)})\rho k_{F}^{4}+\frac{1}{10}t_{1}^{(6)}(1+x_{1}^{(6)})\rho k_{F}^{6}\,, \nn
\end{eqnarray}
%%%
\begin{eqnarray}
\mathcal{V}(^{1}P_1)&=&\frac{3}{160}t_{2}(1-x_{2})\rho k_{F}^{2}+\frac{9}{560}t_{2}^{(4)}(1-x_{2}^{(4)})\rho k_{F}^{4}+\frac{3}{50}t_{2}^{(6)}(1-x_{2}^{(6)})\rho k_{F}^{6} \,,\nnn
\mathcal{V}(^{3}P_0)&=&\frac{3}{160}t_{2}(1+x_{2})\rho k_{F}^{2}+\frac{9}{560}t_{2}^{(4)}(1+x_{2}^{(4)})\rho k_{F}^{4}+\frac{3}{50}t_{2}^{(6)}(1+x_{2}^{(6)})\rho k_{F}^{6} \nonumber\\
& & -\frac{3}{80} t_{o}\rho k_{F}^{2}-\frac{9}{100}t_{o}^{(4)}\rho k_{F}^{4}-\frac{9}{250}t_{0}^{(6)}\rho k_{F}^{6} + \frac{1}{40}W_{o}\rho k_{F}^{2} \,,\nnn
\mathcal{V}(^{3}P_1)&=&\frac{9}{160}t_{2}(1+x_{2})\rho k_{F}^{2}+\frac{27}{560}t_{2}^{(4)}(1+x_{2}^{(4)})\rho k_{F}^{4}+\frac{9}{50}t_{2}^{(6)}(1+x_{2}^{(6)})\rho k_{F}^{6} \nonumber\\
& &+ \frac{9}{160}t_{o}\rho k_{F}^{2}+\frac{27}{200}t_{o}^{(4)}\rho k_{F}^{4}+\frac{27}{500}t_{0}^{(6)}\rho k_{F}^{6} + \frac{3}{80}W_{o}\rho k_{F}^{2} \,,\nnn
\mathcal{V}(^{3}P_2)&=&\frac{3}{32}t_{2}(1+x_{2})\rho k_{F}^{2}+\frac{9}{112}t_{2}^{(4)}(1+x_{2}^{(4)})\rho k_{F}^{4}+\frac{3}{10}t_{2}^{(6)}(1+x_{2}^{(6)})\rho k_{F}^{6} \nonumber\\
& &-\frac{3}{160}t_{o}\rho k_{F}^{2}-\frac{9}{200}t_{o}^{(4)}\rho k_{F}^{4}-\frac{9}{500}t_{0}^{(6)}\rho k_{F}^{6} -\frac{1}{16}W_{o}\rho k_{F}^{2} \,, \nn
\end{eqnarray}
%%%
\begin{eqnarray}
\mathcal{V}(^{1}D_2)&=&\frac{9}{560}t_{1}^{(4)}(1-x_{1}^{(4)})\rho k_{F}^{4}+\frac{1}{10}t_{1}^{(6)}(1-x_{1}^{(6)})\rho k_{F}^{6}\,,\nnn
\mathcal{V}(^{3}D_1)&=&\frac{9}{2800}t_{1}^{(4)}(1+x_{1}^{(4)})\rho k_{F}^{4}+\frac{1}{50}t_{1}^{(6)}(1+x_{1}^{(6)})\rho k_{F}^{6}-\frac{9}{1000}t_{e}^{(4)}\rho k_{F}^{4}-\frac{3}{500}t_{e}^{(6)}\rho k_{F}^{6}\,,\nnn
\mathcal{V}(^{3}D_2)&=&\frac{3}{560}t_{1}^{(4)}(1+x_{1}^{(4)})\rho k_{F}^{4}+\frac{1}{30}t_{1}^{(6)}(1+x_{1}^{(6)})\rho k_{F}^{6} + \frac{3}{200}t_{e}^{(4)}\rho k_{F}^{4}+\frac{1}{100}t_{e}^{(6)}\rho k_{F}^{6}\,,\nnn
\mathcal{V}(^{3}D_3)&=&\frac{3}{400}t_{1}^{(4)}(1+x_{1}^{(4)})\rho k_{F}^{4}+\frac{7}{150}t_{1}^{(6)}(1+x_{1}^{(6)})\rho k_{F}^{6} -\frac{3}{500}t_{e}^{(4)}\rho k_{F}^{4}-\frac{1}{250}t_{e}^{(6)}\rho k_{F}^{6} \,, \nn
\end{eqnarray}
%%%
\begin{eqnarray}
\mathcal{V}(^{1}F_3)&=&\frac{1}{150}t_{2}^{(6)}(1-x_{2}^{(6)})\rho k_{F}^{6}\,,\nnn
\mathcal{V}(^{3}F_2)&=&\frac{1}{70}t_{2}^{(6)}(1+x_{2}^{(6)})\rho k_{F}^{6} -\frac{3}{875}t_{0}^{(6)}\rho k_{F}^{6}\,,\nnn
\mathcal{V}(^{3}F_3)&=&\frac{1}{50}t_{2}^{(6)}(1+x_{2}^{(6)})\rho k_{F}^{6} + \frac{3}{500}t_{0}^{(6)}\rho k_{F}^{6}\,,\nnn
\mathcal{V}(^{3}F_4)&=&\frac{9}{350}t_{2}^{(6)}(1+x_{2}^{(6)})\rho k_{F}^{6} -\frac{9}{3500}t_{0}^{(6)}\rho k_{F}^{6}\,. \nn
\end{eqnarray}
%%%
Up to now, the only available parametrizations of the N3LO pseudo-potentials have been constructed using the $(S,T)$-decomposition and the partial waves analysis (see Ref~\cite{Dav15}).

\bibliographystyle{elsarticle-num}
\bibliography{biblio}

%%%%%%%%%%%%%%%%%%%%%%%%%%%%%%%%%%%%%%%%%%%%

\end{document}